\documentclass[useAMS,usenatbib]{mn2e}
\usepackage{amsmath,amssymb,graphicx,subfigure,rotating,makeidx,changepage}
\usepackage{longtable,lscape,multirow,multicol,calc,pgf,fp,natbib,verbatim}
\usepackage{calc,pgf,fp,natbib,verbatim,lastpage,booktabs}
\usepackage[colorlinks=true,citecolor=blue]{hyperref}
\usepackage[flushleft]{threeparttable}
\usepackage{inputenc,lastpage,booktabs}
\def\deg{\hbox{$^\circ$}}              % degree
\def\arcm{\hbox{$^\prime$}}            % arcminutes
\def\arcmin{\hbox{$^\prime$}}            % arcminutes
\def\zsun{$\rm{Z}_{\odot}$}           %M_sun
\def\td{$\tau_{d}$}
\def\tr{$\tau_{r}$}
\def\swift{{\it Swift}}
\def\kepler{{\it Kepler}}
\def\rosat{{\it ROSAT}}
\def\wasp{SuperWASP}
\def\hipp{{\it Hipparcos}}
\def\asas{ASAS}

\newcommand{\E}[1]{$\times~10^{#1}$}
\newcommand{\Pten}[1]{$10^{#1}$}

\DeclareRobustCommand{\ion}[2]{%
\relax\ifmmode
\ifx\testbx\f@series
{\mathbf{#1\,\mathsc{#2}}}\else
{\mathrm{#1\,\mathsc{#2}}}\fi
\else\textup{#1\,{\mdseries\textsc{#2}}}%
\fi}

\newcommand{\changecodes}{%
  \count255=`A
  \loop
  \mathcode\count255=\numexpr\mathcode\count255-\string"100\relax
  \ifnum\count255<`Z
    \advance\count255 1
  \repeat
  \count255=`a
  \loop
  \mathcode\count255=\numexpr\mathcode\count255-\string"100\relax
  \ifnum\count255<`z
    \advance\count255 1
  \repeat
}
\bibpunct{(}{)}{;}{a}{,}{,}

\title[Active ultrafast rotator LO Peg]{LO Peg: surface differential rotation, flares, and spot-topographic evolution}

 \author[Karmakar et al.]{Subhajeet Karmakar$^{1}$\thanks{E-mail:
 subhajeet@aries.res.in}, J. C. Pandey$^{1}$, I. S. Savanov$^{2}$, G. Ta\c{s}$^{3}$, S. B. Pandey$^{1}$, \newauthor K. Misra$^{1}$, S. Joshi$^{1}$, E. S. Dmitrienko$^4$, T. Sakamoto$^{5}$, N. Gehrels$^{6}$,  and \newauthor T. Okajima$^{6}$\\
 $^{1}$ Aryabhatta Research Institute of observational sciencES (ARIES), Manora Peak, Nainital 263002, India\\
 $^{2}$ Institute of Astronomy, Russian Academy of Sciences, ul. Pyatniskaya 48, Moscow 119017, Russia\\
 $^{3}$ Astronomy and Space Sciences Department, Science Faculty, Ege University, \.{I}zmir, Turkey\\
 $^{4}$ Sternberg Astronomical Institute, Moscow State University, Universiteskii pr 13, Moscow, 119992 Russia\\
 $^{5}$ Department of Physics and Mathematics, Aoyama Gakuin University 5-10-1 Fuchinobe, Chuo-ku Sagamihara-shi\\ Kanagawa 252-5258, Japan\\
 $^{6}$ NASA Goddard Space Flight Center Greenbelt, MD 20771, USA
} 

\begin{document}

\date{Accepted 2016 April 11. Received 2016 April 11; in original form 2015 September 12}
 \pagerange{\pageref{firstpage}--\pageref{LastPage}} \pubyear{2016}
 \maketitle
% \makeindex
 \label{firstpage}

%%%%%%%%%% ABSTRACT %%%%%%%%%%%%%%%%%%%%%%%%%%%%%%%%%%%%%%%%%%%%%%%%%%%%%%%%%%%%%%
\begin{abstract}
Using the wealth of $\sim$24 yr multiband data, we present an in-depth study of the star-spot cycles, surface differential rotations (SDR), optical flares, evolution of star-spot distributions, and coronal activities on the surface of young, single, main-sequence, ultrafast rotator (UFR) LO Peg. From the long-term $V$-band photometry, we derive rotational period of LO Peg to be 0.4231 $\pm$ 0.0001 d. Using the seasonal variations on the rotational period, the SDR pattern is investigated, and shows a solar-like pattern of SDR. A cyclic pattern with period of $\sim$2.7 yr appears to be present in rotational period variation. During the observations, 20 optical flares are detected with a flare frequency of $\sim$1 flare per two days and with flare energy  of $\sim$\Pten{31-34} ~erg. The surface coverage of cool spots is found to be in the range of $\sim$9--26 per cent. It appears that the high- and low-latitude spots are interchanging their positions. Quasi-simultaneous observations in X-ray, UV, and optical photometric bands show a signature of an excess of X-ray and UV activities in spotted regions.
\end{abstract}

%%%%%%%%%% SECTION 1 %%%%%%%%%%%%%%%%%%%%%%%%%%%%%%%%%%%%%%%%%%%%%%%%%%%%%%%%%%%%%
\section[]{Introduction}     
\label{sec:intro}
Stars with spectral type from late-F to early-K have a convective envelope above a radiative interior with an interface where strong shear leads to amplification of magnetic fields. Observations of these stars provide good constraints on present theoretical dynamo models, which are developed on the basis of the Sun. The solar activity cycle is believed to be generated through dynamo mechanism operating either in the convection zone or in the stably stratified layer beneath it. Stars with a similar internal structure to that of the Sun are also expected to show the solar-type dynamo operation. 
The strong dynamo in solar-type stars leads to rich variety of magnetic activities such as surface inhomogeneities due to the presence of dark spots, short- and long-term variations in spot cycles, and flares.

Dark spots move across the stellar disc due to the stellar rotation and thus modulate the
total brightness with the rotational period of the star which in turn allows us to derive the stellar rotational period. 
The spots on the stellar surface have been imaged by using a variety of techniques like Doppler imaging \citep{Vogt-83-5} and interferometric technique \citep{Parks-11-1}. However, high-resolution spectroscopic observations with a high-signal-to-noise ratio and a good phase coverage as required for Doppler imaging are limited. Further, the Doppler imaging technique can only be applied for fast-rotating stars with low inclination, whereas, the interferometric technique can be used for nearby stars of large angular size. The vast majority of spotted stars cannot be imaged with either of these techniques. Therefore, long-term traditional photometric observations are important to understand the active region evolution and the stellar activity cycles \citep[e.g.][]{Jarvinen-05-2, Olah-09, Roettenbacher-13-1}. 
Since a light curve represents a one-dimensional time series, the resulting stellar image contains information mostly in the direction of rotation, i.e., in the longitude, rather than spot size and locations in the latitude \citep[][]{Savanov-08-2}. Although the projection effects and limb darkening allow the inversion technique to recover more structures than is obvious at first glance. 
The  surface differential rotations (SDR) in these stars can be determined by measuring the rotational period of stars with spots. Several authors in the past have studied SDR in solar-type stars using long-term photometry \citep[e.g.][]{Baliunas-85-9, Hall-91-15, Walker-07-27,Reinhold-13-1,Reinhold-15-2}. Since spots cover a limited range of latitudes on the stellar surface; therefore,  amplitudes of SDR derived with this method give the lower limits. The season-to-season variations of the rotational period as measured from spectrophotometric \citep[][]{Donahue-96-11} or broad-band photometric observations can be termed as a proxy of stellar butterfly diagram. In analogy with the Sun, such diagrams are interpreted in terms of migration of activity centres towards latitudes with different angular velocities. 
Another consequence of stellar magnetic activities are flares, which are the result of reconnection of magnetic field lines at coronal height. Flares are explosions on the stellar surface releasing huge amount of the magnetic energy stored near star-spots in the outer atmosphere of stars \citep[e.g.][]{Gershberg-05, Benz-10-2, Shibata-11-1}. Observationally flares are detected over all frequencies of the electro-magnetic spectrum. The average flare duration is \Pten{2-4} s \citep{Kuijpers-89}. The total energy released during a flare (in all wavelengths) is \Pten{34-36} erg, i.e. \Pten{2-4} times more powerful that the solar analogue \citep{Byrne-89-1}.

In this paper, we have investigated an active, young, single, main-sequence, K5--8 type ultrafast rotator (UFR) LO Peg.
LO Peg has been an interesting object to study over the last two decades.
From photometric observations, \cite{Barnes-05-32} derived a rotational period of 0.42323 d.  
A presence of strong flaring activity was also identified by \cite{Jeffries-94-2} and \cite{Eibe-99} from H$\alpha$ and He{\sc i} D3 observations. \cite{Tas-11} found evidence of flares in the optical band. Doppler imaging of LO Peg showed evidence of high polar activities \citep{Lister-99-3, Barnes-05-32, Piluso-08}. 
Several photometric, polarimetric, and X-ray studies were also carried out by \cite{Dal-03, Pandey-05-8, Pandey-09-11, Csorvasi-06}, and \cite{Tas-11}. 
 The above results encouraged us to collect all available data and analyse them with the aim to establish whether the star exhibits active longitudes and cyclic behaviour in spot patterns and overall spot activity. 

 The paper is organized as follows: in \S \ref{sec:obs} we provide details on observational data sets and discuss the data analysis techniques. In \S \ref{sec:analysis-results}, we present our analysis and results on light curves, SDR, flaring activity, surface inhomogeneity, and coronal activities.
 Finally, we discuss all the results in the light of present understanding in \S \ref{sec:discussion} and a brief summary of our results is given in \S \ref{sec:summary}.

%%%%%%%%%% SECTION 2 %%%%%%%%%%%%%%%%%%%%%%%%%%%%%%%%%%%%%%%%%%%%%%%%%%%%%%%%%%%%%
\section[]{Observations and data reduction}
\label{sec:obs}
%********** fig :1: Landolt-WASP Calibration ********************
%\begin{uprightmath}
\begin{figure}
\centering
\includegraphics[width=6.2cm,angle=-90]{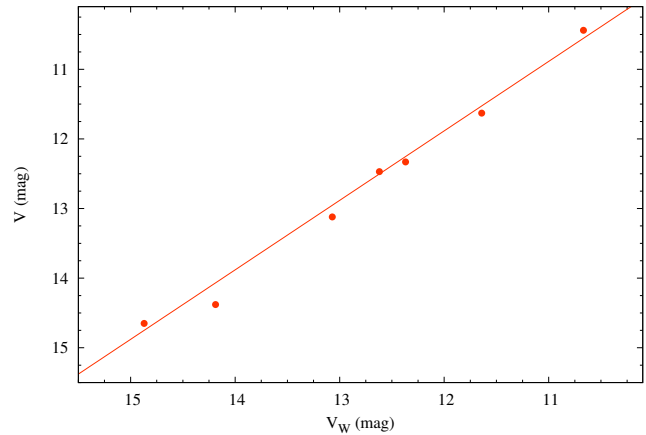}

\caption{The relation between \wasp ~magnitude ($V_{\rm W}$) and  Johnson $V$-magnitude (V) of  Landolt standard  stars (TPHE field). Error bars shown in both axes are less than the size of the symbol. Continuous line shows the best--fitting straight line. We derived the relation between $V$ and $V_{\rm W}$ as $ V = V_{\rm W} - 0.09$.}
\label{fig:WASP_calib}
\end{figure}
%\end{uprightmath}

%****************************************************************
\subsection{Optical data}
\label{subsec:obs_optical}
We observed LO Peg on 30 nights between 2009 October 25 and 2013 December 18 in Johnson $U$, $B$, $V$, and $R$ photometric bands with the 2-m IUCAA Girawali Observatory \citep[IGO; see][]{Das-99-37}, 1.04-m ARIES Sampurnanand Telescope \citep[ST; see][]{Sinvhal-72} and  0.36-m Goddard Robotic Telescope \citep[GRT; see][]{Sakamoto-09-12}. The exposure time was between 5 and 60 s depending on the seeing condition,  filter, and telescope used. Several bias and  twilight flat frames were taken in each observing night. Bias subtraction, flat-fielding, and aperture photometry were performed using the standard tasks in {\sc IRAF}\footnote{http://iraf.net}.
In order to get the standard magnitude of the program star,  differential photometry had been adopted, assuming that the errors introduced due to colour differences between comparison and program stars are very much small. We have chosen TYC 2188-1288-1 and TYC 2188-700-1  as the comparison and check stars, respectively. The differences in the measured $U$, $B$, $V$, and $R$ magnitudes of comparison and check stars did not show any secular trend during our observations. The nightly means of standard deviations of these differences were 0.009, 0.008, 0.008, and 0.007 mag in $U$, $B$, $V$, and $R$ bands, respectively. This  indicates that both the comparison and check stars were constant during the observing run. 
 The standard magnitudes of comparison and check stars were taken from NOMAD Catalogue \citep{Zacharias-04-1}. The derived photometric uncertainties for program star, check star, and comparison star were propagated to get the final photometric uncertainty of LO Peg. 

%********** tab :1: Optical Log *********************************
\begin{table*}
\scriptsize
\tabcolsep=0.53cm
\begin{adjustwidth}{0cm}{}
\caption{Log of optical observations of LO Peg.}
\label{tab_optical-log}
\begin{threeparttable}
\begin{tabular}{llllllll}
\hline\hline

\multirow{2}{*}{\textbf{Observatories}} 	   &\textbf{Start HJD}       & \textbf{End HJD}  	  &\multicolumn{4}{c}{\textbf{Number of exposures}}   & \multirow{2}{*}{\textbf{Ref}}        \\        
\cline{4-7}
&\textbf{(2400000+)}     &  \textbf{(2400000+)}         &\textbf{$U$-band}  & \textbf{$B$-band}   &\textbf{$V$-band}   &\textbf{$R$-band}      &         \\        
 \hline\hline

ARIES ST  & 55130.118      & 56645.365  &    5  &       67   &      72    &      5  &  P          \\         
GRT	  & 55766.811      & 55775.771  &  ---  &      4     &      9     &     3   &  P           \\
IGO       & 55130.100      & 55135.130  &  ---  &      10    &     30     &   ---   &  P           \\            
\hline
Archive  \\ 
\hline
\hipp     & 47857.501      & 48972.277  &  ---  &      ---   &    136$^\dagger$    &  ---    &  a 	           \\
\asas     & 52755.911      & 55092.673  &  ---  &      ---   &    259         &   ---   &  b                  \\     
\wasp     & 53128.655      & 54410.482  &  ---  &    ---     & 8047$^\dagger$&         &  c  		   \\    	    
\hline
Literature                                                                                                     \\
\hline
JKT 	  & 48874.519      & 48883.592  & 25	  &     25     &     25	    &   25    &  d    \\                
ARIES ST  & 52181.167      & 54421.047  &  ---  &      90    &     119    &    ---  &  e  \\       
EUO       & 52851.445      & 55071.521  & 5566  &      5566  &     5566   &   5566  &  f    \\   	    

\hline\hline
\end{tabular}
    \begin{tablenotes}
  \item \textbf{Notes. }
  \item 
P -- Present study; a -- \hipp ~archive ; b -- \asas ~archive; c -- \wasp ~archive 
  \item 
d -- \cite{Jeffries-94-2}; e -- \cite{Pandey-05-8,Pandey-09-11}; f -- \cite{Tas-11}. 
  \item 
$\dagger$ - \hipp ~and \wasp ~data were converted to corresponding $V$-band magnitude.     
 \end{tablenotes}
   \end{threeparttable}
\normalsize
\end{adjustwidth}
\end{table*}

%****************************************************************
We have compiled various other available data sets in $U$, $B$, $V$, and $R$ band from literature \citep{Jeffries-94-2, Pandey-05-8, Pandey-09-11, Tas-11} and from archives to supplement our data sets. The archival data were taken from \hipp\footnote{http://heasarc.gsfc.nasa.gov/W3Browse/all/hipparcos.html} \citep{Perryman-97-6}, All Sky Automated Survey\footnote{http://www.astrouw.edu.pl/asas/?page=main} \citep[\asas;][]{Pojmansky-02}, and Super Wide Angle Search for Planets\footnote{http://exoplanetarchive.ipac.caltech.edu/applications/TblSearch/\\tblSearch.html?app=ExoSearch\&config=superwasptimeseries} \citep[\wasp;][]{Pollacco-06-1} observations. 
The log of optical observations is listed in Table \ref{tab_optical-log}. \hipp ~observations spanned over $\sim$3 yr from 1989 November 27 to 1992 December 15. The \hipp ~magnitude ($V_{\rm H}$) was converted to Johnson V magnitude by using the  relation
$V = V_{\rm H} - (V-I)_{c}$, where $(V-I)_{c}$ is the catalogue value corresponding to the colour ($V-I$). With a ($V-I$) colour of 1.288 mag for G-M dwarfs, we get the $(V-I)_{c}$ value of LO Peg to be 0.124 mag. 
\asas ~survey was done in $V$-band and has a much longer observing span of $\sim$7 yr (2003 April 26--2009 October 1). In the \asas ~observations,  we have used only `A' and `B' grade data within 1\arcsec ~of the star LO Peg. \asas ~photometry provides five sets of magnitudes corresponding to five aperture values varying in size from 2 to 6 pixels in diameter. For bright objects, \cite{Pojmansky-02} suggested that these magnitudes corresponding to the largest aperture (diameter 6 pixels) are useful. Therefore, we took magnitudes corresponding to the largest aperture for further analysis.
\wasp ~observations of LO Peg during 2004 May 3 to 2006 June were unfiltered which were not useful for our study \citep[see][]{Pollacco-06-1}. A broad-band filter with a pass-band from 400 to 700 nm (known as \wasp ~$V$-band) was installed on 2006 June. In our analysis,  we make use of the data taken from 2006 June, onwards. Since the \wasp ~data were taken in a broader band than the Johnson $V$-band; it is necessary  to convert \wasp ~band magnitude ($V_{\rm W}$) to Johnson $V$ magnitude. Fortunately, the Landolt standard field TPHE with seven standard stars was observed by \wasp. Fig. \ref{fig:WASP_calib} shows the plot between $V$ and $V_{W}$  of  Landolt standard  stars, where the continuous line shows the best fit straight line.   We derived the relation between $V$ and $V_{\rm W}$ as $ V = V_{\rm W} - 0.09$, and converted the $V_{\rm W}$ magnitude into $V$.  Further, we have restricted our analysis  within magnitude error less than or equal to 0.04 mag both in \asas ~and \wasp ~data.
Including present observations along with the data compiled from literature and archive, LO Peg was observed for $\sim$24.1 yr from 1989 to 2013.

\subsection{X-ray and UV data}
LO Peg was observed in 17 epochs with \swift ~satellite (P.I. Pandey, ID: 0123720201) from 2008 April 30 to 2012 July 2. The observations were made in soft X-ray band (0.3--10.0 keV) with X-Ray Telescope \citep[XRT; ][]{Burrows-05-3} in conjunction with  UV/Optical Telescope  \citep[UVOT; ][]{Gehrels-04-5} in UV bands (170--650 nm). The offset of the observations lies between 1\arcm.11 and 4\arcmin.35. The XRT exposure time of LO Peg ranges from 0.3 ks to 5.0 ks. X-ray light curves and spectra of LO Peg were extracted from on-source counts obtained from a circular region of 36\arcsec ~on the sky centered on the X-ray peaks. Whereas, the  background was extracted from an annular region having an inner circle of 75\arcsec ~and outer circle of 400\arcsec ~co-axially centered on the X-ray peaks.  The X-ray light curves and spectra  for the source and background were extracted using the {\sc xselect} package. In order to see the long-term X-ray variation we converted \rosat ~~Position Sensitive Proportional Counter (PSPC) count rate \citep{Pandey-05-8} to \swift ~XRT count rate with multiple component models of {\sc webpimms} (see \S ~\ref{subsec:x-ray} for further details). Simultaneous observations of LO Peg with \swift ~UVOT were carried out in $uvw$2 (192.8 nm), $uvm$2 (224.6 nm) and $uvw$1 (260.0 nm) filters \citep{Roming-05-2} with exposure times  between 0.02 and 3.06 ks. UV light curves were extracted using the {\sc uvotmaghist} task.

%%%%%%%%%% SECTION 3 %%%%%%%%%%%%%%%%%%%%%%%%%%%%%%%%%%%%%%%%%%%%%%%%%%%%%%%%%%%%%
\section{Analysis and Results}
\label{sec:analysis-results}
%********* fig :2: Long term Multi-band Light curve *************

\begin{figure*}
%{multiband_lc_lopeg_plot.pdf}
\includegraphics[width=17cm,angle=0]
{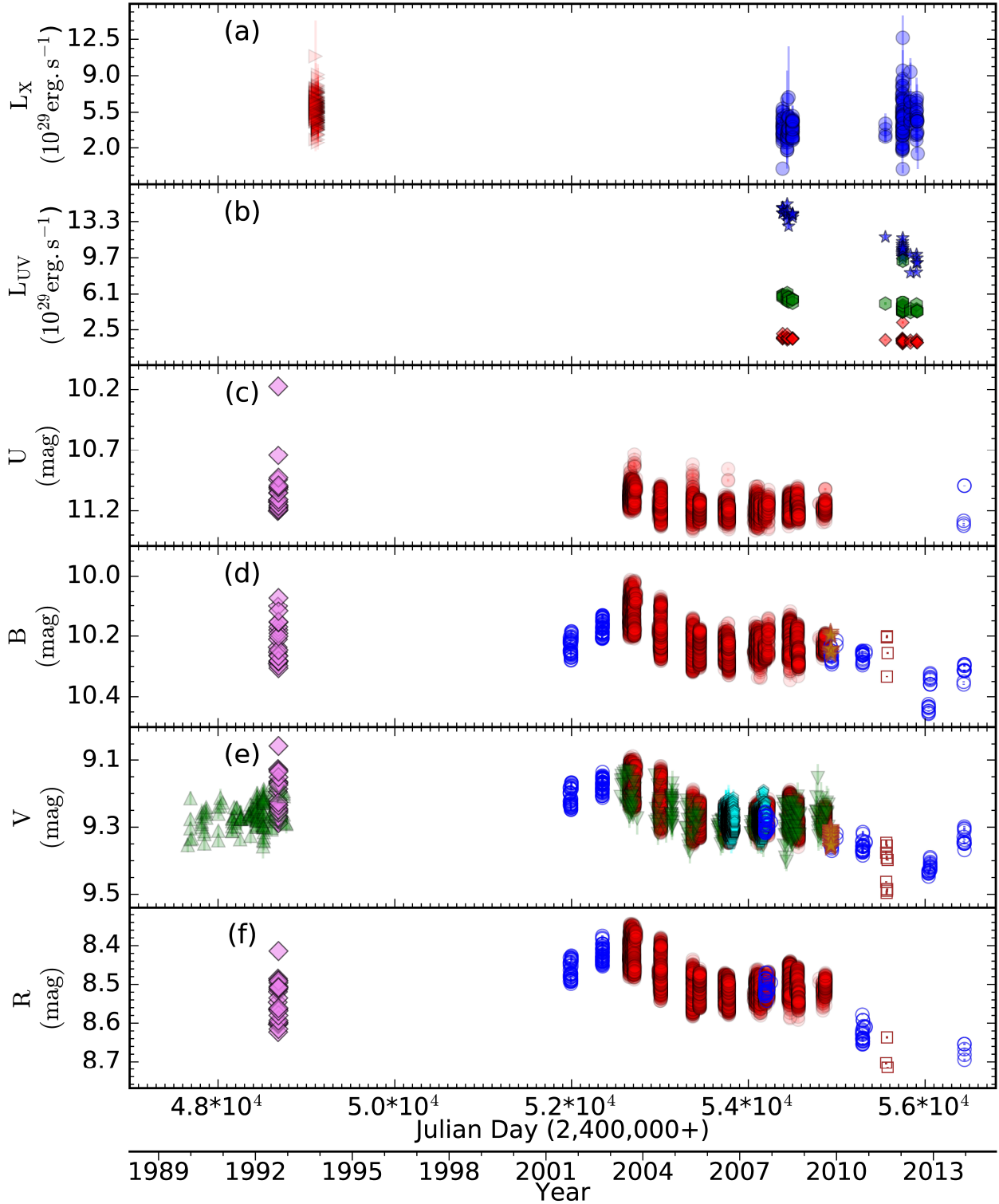}   %this is the final plot in eps with best resoluton
%--------------------------------------------------------------------------------
\caption{Multiband light curve of LO Peg. From top to bottom -- (a) the X-ray light curves obtained from \swift ~XRT (solid circle) and \rosat ~PSPC (solid right triangle) instruments. (b) The UV light curve obtained from \swift ~UVOT in three different UV-filters: $uvw$2 (solid hexagon), $uvm$2 (solid diamond), and $uvw$1 (solid star). (c--f) The next four panels shows optical light curves obtained in  $U$, $B$, $V$, and $R$  bands, respectively. Observations were taken from ARIES (open circle), IGO (solid star), GRT (open square), EUO (solid circle), and JKT (solid diamond) telescopes and archival data were obtained from \hipp ~satellite (solid triangle), \wasp ~(solid pentagon), and \asas ~(solid reverse triangle).}
\label{fig_multiband-lc}
\end{figure*}

%****************************************************************
\subsection{Optical light curves and period analysis}
\label{subsec:lc-period}
Fig. \ref{fig_multiband-lc} shows the multiwavelength light curves  of LO Peg where bottom four panels indicate the optical $U$, $B$, $V$, and $R$ photometric bands. The optical light curves display high amplitude both in short-term and long-term flux variations. 
The most populous $V$-band data  was analysed  for the periodicity using Scargle--Press period search method \citep{Scargle-82, Horne-86-5, Press-89-3} available in the UK Starlink {\sc period} package \citep[version-5.0-2; see][]{Dhillon-01-2}.    
%*********** fig :3: Fourier Transform***************************
\begin{figure*}
\centering
\subfigure[]{\includegraphics[width=8.9cm,angle=0]{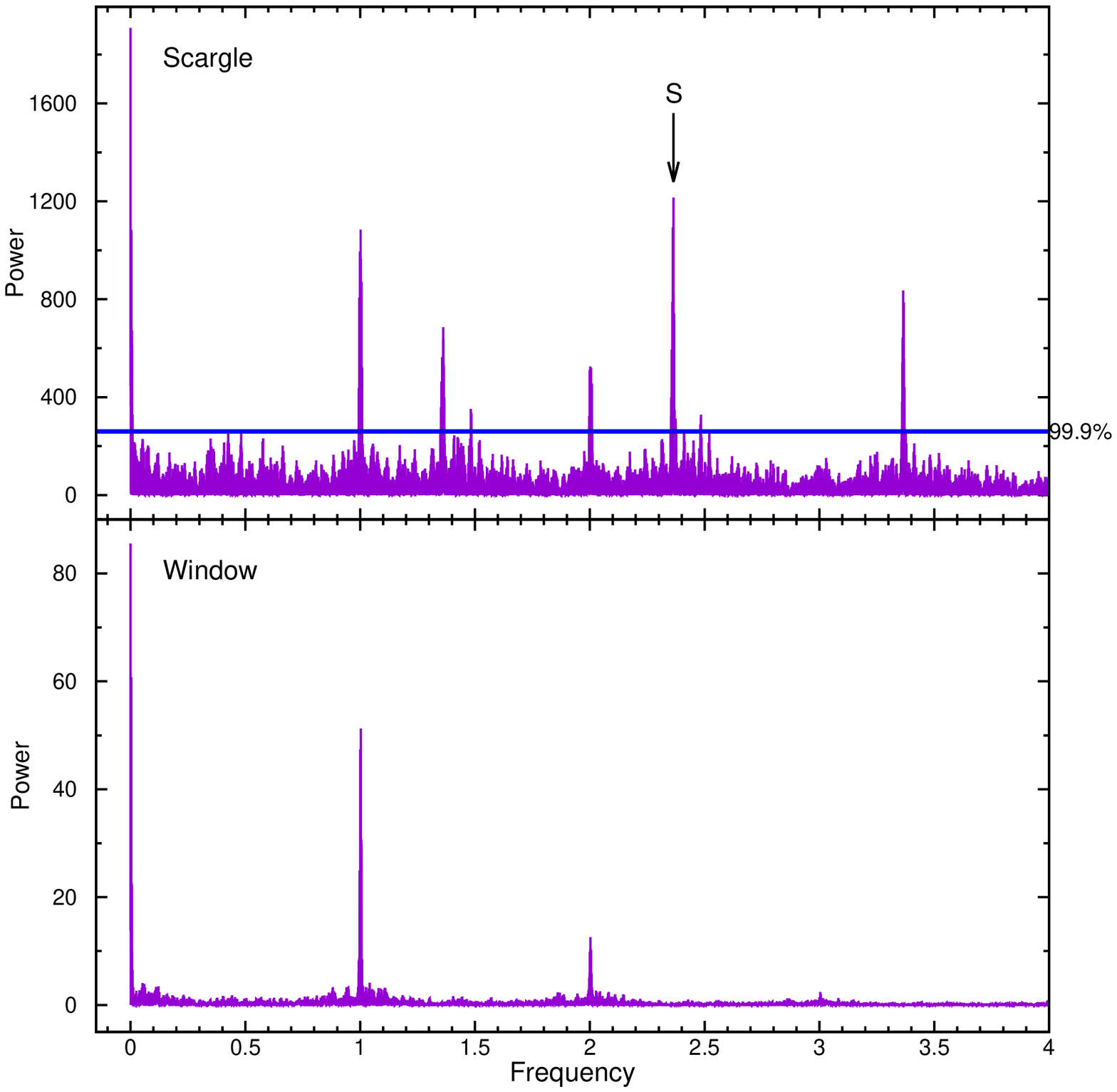}}
\subfigure[]{\includegraphics[width=7.1cm,angle=0]{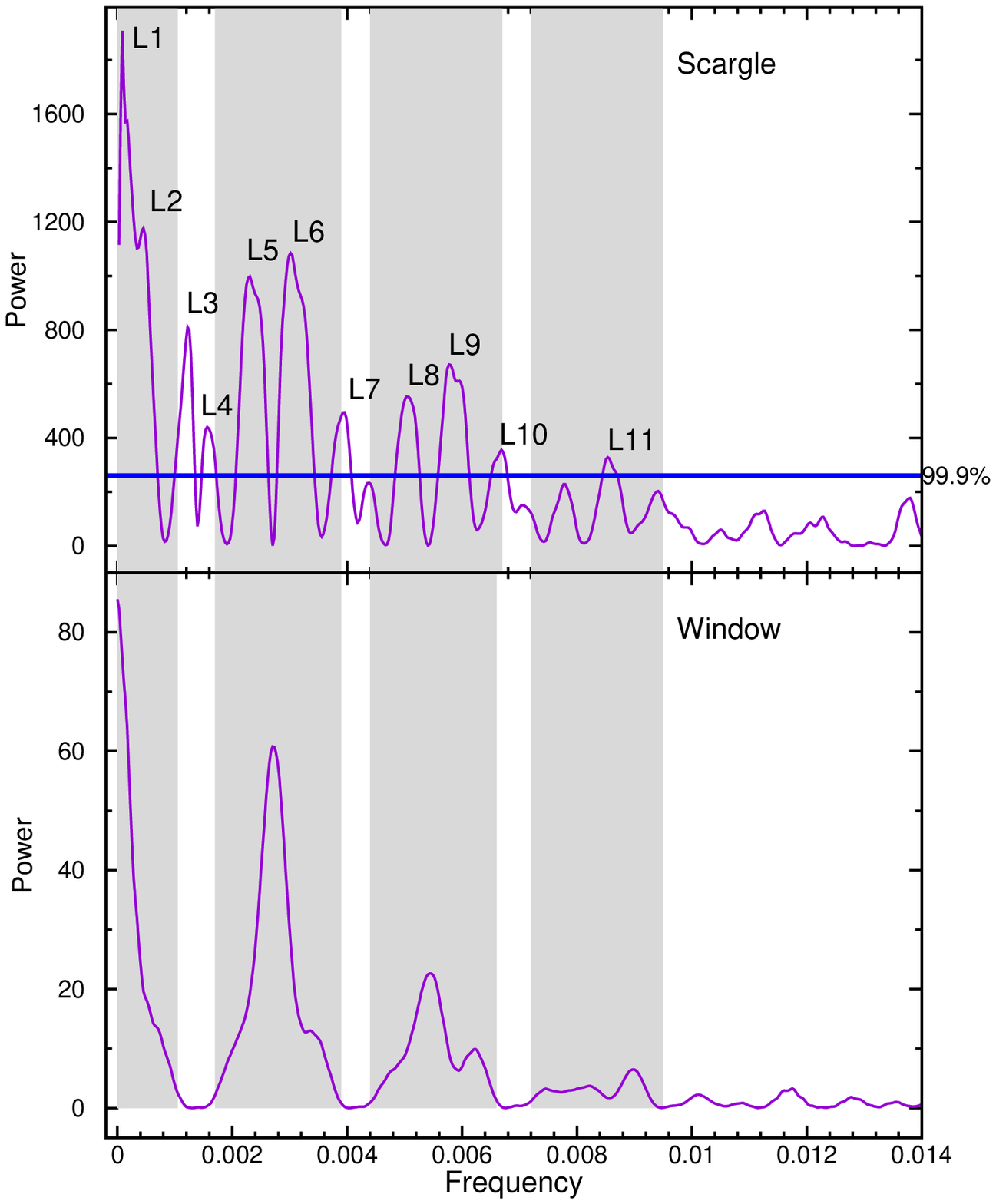}}
\caption{(a) Scargle-Press periodogram obtained from $V$-band data (top) along with the calculated window function (bottom). The significance level of 99.9\% is shown by continuous horizontal line. The peak marked as `S' corresponds to the stellar rotation period. (b) The low frequency region is zoomed to show the long term periods, the shaded regions show the frequency domain ascribed to window function. Long-term peaks are marked by `L1--L11' (see the text for detailed description).}

\label{fig_ft}
\end{figure*}

%****************************************************************
Top panel of Fig. \ref{fig_ft}(a) shows the power spectra obtained from Scargle periodogram.
We have also calculated the False Alarm Probability (FAP) for any peak frequency using the method given by \cite{Horne-86-5}. The significance level of 99.9\%  is shown by continuous horizontal line in the Fig. \ref{fig_ft}. 
Large and almost periodic gaps in the data set led to further complications in the power spectrum. True frequencies of the source were further modulated by the irregular infrequent sampling defined by window function of the data. In order to resolve this problem, we have computed window function with the same time sampling and photometric errors of the actual light curve, but contains only a constant magnitude as the average magnitude of the data (9.250 mag). We have repeated the process with many realization of noise, where we have generated 1000 random numbers within 3$\sigma$ range of the mean value and taking these value as a constant we computed each periodogram. The resulting periodograms were averaged and shown in bottom panel of Fig. \ref{fig_ft}.

%*********** fig :4: Fold at long-term **************************
\begin{figure}
\centering
\includegraphics[height=9.15cm,angle=0]{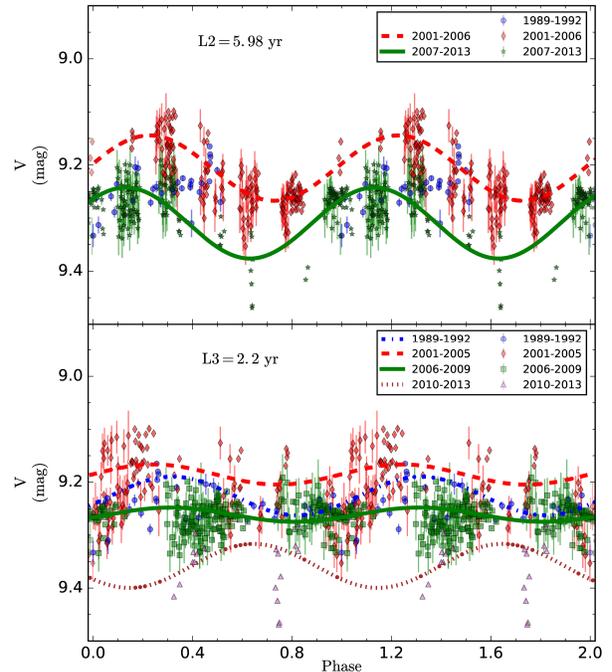}
\caption{Folded light curve of LO Peg in $V$-band with periods of 5.98 yr (top) and 2.2 yr (bottom). The best-fitting sinusoids are shown by dashed, dotted, dash--dotted, and continuous lines for different time-intervals marked at the top-right corner of each panel. We could not fit the time-interval of 1989--1992 in top panel due to partial phase coverage of data points.}
\label{fig_folded}
\end{figure}

%****************************************************************
In Scargle power spectra the peak marked as `S'  corresponds to the rotational period of 0.422923 $\pm$ 0.000005 d, where the uncertainty in period was derived using the method given by \cite{Horne-86-5}. The uncertainty in the derived period was very small ($< 1$ s) due to the long base line of unevenly sample  data. Further, there was a large gap in the data. Therefore, we derived the rotational period and corresponding error by averaging the seasonal rotational periods (see \S \ref{subsec:diff_rot}). The value of mean  seasonal rotation period was found to be 0.4231 $\pm$ 0.0001 d.  
Two other smaller peaks at periods of $\sim 0.212 \rm~{d}$ and  $\sim 0.846 \rm~{d}$  were identified in the power spectra as the harmonic and sub-harmonic, respectively. The former may indicate the existence of two active regions over the surface of LO Peg, whereas the later appears due to repeated occurrence of the same spot at multiple of it's period. In order to search for the long term periodicity, we have zoomed the lower frequency range  of  Fig. \ref{fig_ft}(a)  and shown in Fig. \ref{fig_ft}(b). Several peaks were found above the 99.9\% confidence level, these peaks are marked by L$i$, where $i$ = 1 to 11. However, many of the peaks were found under window function (see shaded region of  Fig. \ref{fig_ft}b). Peaks corresponding to the periods L3 and L4 did not fall under the window function. Further, we have folded the data in each period and found periodic modulation only for periods 5.98 and 2.2 yr corresponding to L2 and L3. To avoid the modulation due to its rotation,
we have made one point of every five-rotation period ($\sim 2$ d). Further, the light curve evolved over a long time; therefore, for folding the  data on the long periods, we have split the light curves for different time segments such that each segment has a length of minimum to those long periods. The top and bottom panels of Fig \ref{fig_folded} show the folded light curves on periods 5.98 and 2.2 yr for different time segments, respectively. For the period 5.98 yr, we found only three time segments of $\sim$6 yr. For 2.2 yr period, we could make only four time-segments of each 3-4 yr. 
 Further, we have fitted sinusoids in the phase folded light curves for each interval. In the top panel of Fig. \ref{fig_folded}, the best-fitting sinusoids are shown by dashed and continuous curves for the time interval 2001--2006 and 2007--2013, respectively. Sinusoid was not fitted to the time interval of 1989--1992 due to the partial phase coverage. The long-term evolution of the activity seems to present.   Similar behaviour of the light curve was also seen while fitting sinusoid to the phase folded light curves on 2.2 yr in the bottom panel of Fig. \ref{fig_folded}.
%%%%%%%%%% SECTION 4 %%%%%%%%%%%%%%%%%%%%%%%%%%%%%%%%%%%%%%%%%%%%%%%%%%%%%%%%%%%%%
%********** tab :2: Differential Rotation ***********************
\begin{table*}
  \scriptsize
  \tabcolsep=0.38cm
  \begin{adjustwidth}{-0cm}{}
  \caption{Parameters derived from SDR analysis.}
  \label{tab_diff-rot}
\begin{threeparttable}
     \begin{tabular}{cccccccc}
       \hline\hline
       \multirow{2}{*}{{\bf Cycle}$^{a}$}&	{\bf Start HJD}	& 	 {\bf END HJD}	&	 {\bf MEAN HJD}	& \multirow{2}{*}{$\mathbf{{\mathit{N}_{0}}}$}&		$\mathbf{\mathit{V}_{avg}}$			&$\mathbf{\mathit{P}_{sr}}$	& \multirow{2}{*}{{\bf FAP}}  \\ %&\\&	{\bf Slope}				\\ %&\\  
			&	{\bf (2400000+)}& 	 {\bf (2400000+)}&	 {\bf (2400000+)}&               &		\textbf{(mag)}				&	\textbf{(d)} 				& 			  \\ %&\\&	{\bf Slope}				\\ %&\\  
\hline\hline                                                                                                                                                                                                                                             
\multirow{4}{*}{I}	& 47857.500    	& 48066.063 	 & 47961.782	  &	19		& 9.268		$\pm$		0.004 	& 0.4243        $\pm$	0.0003 	& 0.05	      \\ 
			& 48113.175    	& 48368.850      & 48282.378      &	13		& 9.255		$\pm$		0.005 	& 0.4313        $\pm$	0.0005 	& 0.12	      \\ 
			& 48368.850    	& 48624.525      & 48490.262      &	30		& 9.248		$\pm$		0.003 	& 0.4203        $\pm$	0.0002 	& 0.03	      \\                                                                                                                                                                                                                                  
			& 48624.525    	& 48880.200      & 48756.841      &	73		& 9.233		$\pm$		0.002 	& 0.4231        $\pm$	0.0001 	& 3.28e-08    \\
\\ %&\\ 
\multirow{1}{*}{II}	& 48880.200    	& 48972.277      & 48926.410      &	25		& 9.194		$\pm$		0.002 	& 0.4237        $\pm$	0.0004 	& 2.31e-03    \\                                                                                                                                                                                                                                       
\\ %&\\
\multirow{2}{*}{V}	& 52181.166    	& 52198.126      & 52189.646      &	37		& 9.1909	$\pm$		0.001 	& 0.4240        $\pm$	0.0009 	& 1.25e-06    \\ 
			& 52546.207    	& 52551.254      & 52548.731      &	50		& 9.154		$\pm$		0.001 	& 0.421        $\pm$	0.001 	& 1.99e-07    \\                                                                                                                                                                                                                                       
\\ %&\\
\multirow{3}{*}{VI}	& 52755.911    	& 52942.538      & 52849.224      &	894		& 9.1522	$\pm$		0.0003 	& 0.42319        $\pm$	0.00002	& 2.45e-165   \\ 
			& 53142.922    	& 53344.523      & 53243.722      &	574		& 9.2100	$\pm$		0.0005 	& 0.42308        $\pm$	0.00005 & 6.20e-105   \\ 
			& 53487.918    	& 53650.354      & 53569.136      &	889		& 9.2546	$\pm$		0.0003 	& 0.41864        $\pm$	0.00004 & 1.14e-144   \\ 
\\ %&\\
\multirow{3}{*}{VII}	& 53853.920    	& 54044.457      & 53949.188      &	4792		& 9.2617	$\pm$		0.0002 	& 0.42295        $\pm$	0.00001 & $\sim$0      \\                                                                                                                                                                                                                                       
			& 54227.896    	& 54454.050      & 54340.973      &	4700		& 9.2610	$\pm$		0.0002 	& 0.42331        $\pm$	0.00001 & $\sim$0      \\ 
			& 54590.916    	& 54785.517      & 54688.216      &	1061		& 9.2487	$\pm$		0.0004 	& 0.42305        $\pm$	0.00002 & 3.37e-177   \\ 
\\ %&\\
\multirow{2}{*}{VIII}	& 54954.917    	& 55196.045      & 55075.481      &	386		& 9.2603	$\pm$		0.0008 	& 0.4304        $\pm$	0.0001	& 1.48e-09    \\ 
			& 55489.158    	& 55526.101      & 55507.630      &	23		& 9.338		$\pm$		0.001 	& 0.417        $\pm$	0.001 	& 0.05	      \\                                                                                                                                                                                                                                       
\\ %&\\
\multirow{3}{*}{IX}	& 55758.817    	& 55775.770      & 55767.294      &	9		& 9.402		$\pm$		0.003  	& 0.424        $\pm$	0.001 	& 0.22	      \\ 
			& 56239.143    	& 56257.173      & 56248.158      &	20		& 9.402		$\pm$		0.001 	& 0.424        $\pm$	0.001 	& 0.01	      \\ 
			& 56636.357    	& 56645.364      & 56640.861      &	12		& 9.312		$\pm$		0.001 	& 0.419        $\pm$	0.002 	& 0.05	      \\                                                                                                                                                                                                                                       

\hline\hline
     \end{tabular}
    \begin{tablenotes}
  \scriptsize
  %\small
  \item \textbf{Notes. }
  \item 
    $a$ -- Detected star-spot cycles of 2.7$\pm$0.1 yr (shown in bottom panel of Fig. \ref{fig_diff-rot}a), $N_\mathrm{0}$ is the number of data points during each season, $\mathrm{\mathit{V}_{avg}}$ is the average $V$-band magnitude in each season, $P_\mathrm{sr}$ is the seasonal rotational period, and FAP is false alarm probability.
       \end{tablenotes}
  \end{threeparttable}
  \normalsize
  \end{adjustwidth}
\end{table*}

%****************************************************************

\subsection{Surface Differential Rotation}
\label{subsec:diff_rot}
%*********** fig :5: Differential Rotation **********************
\begin{figure*}
\centering
\subfigure[]{\includegraphics[width=8cm,angle=0]{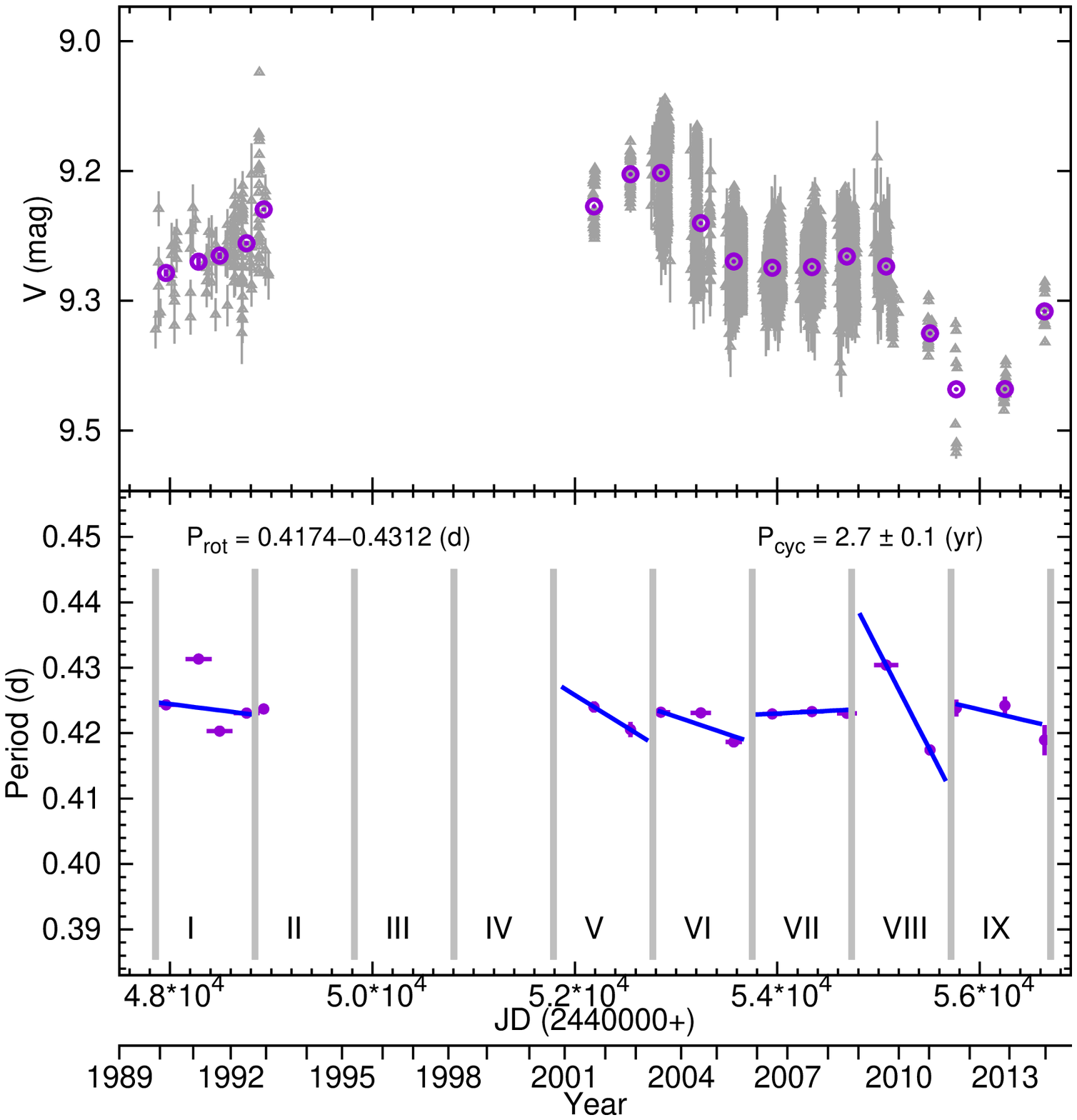}}
\subfigure[]{\includegraphics[width=8cm,angle=0]{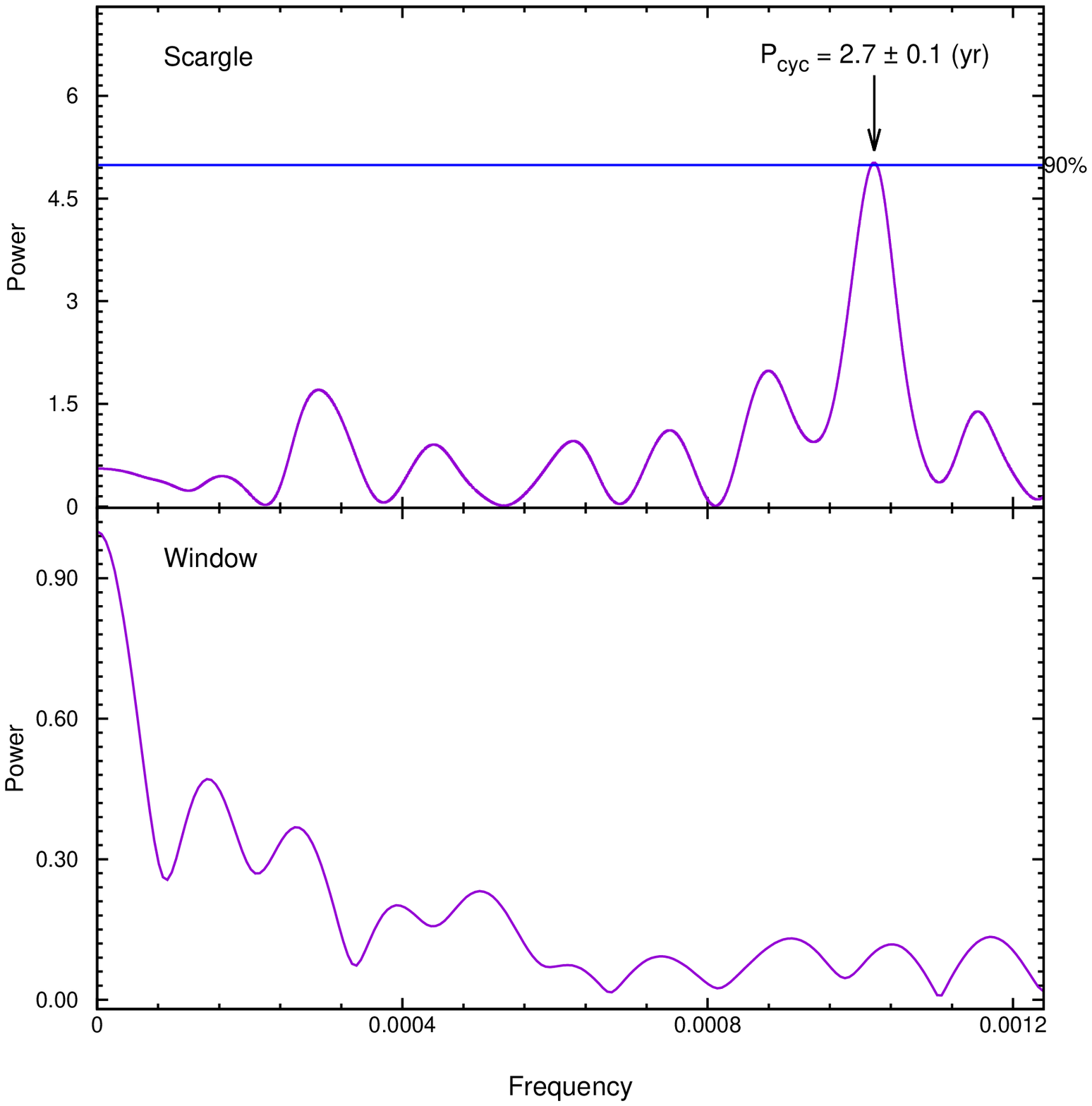}}
\caption{Top panel of (a) shows the $V$-band light curve (solid triangles) along with the mean magnitude of each season (open circles). Solid circles in bottom panel of (a) shows the derived rotational periods in each season. The Scargle-Press periodogram  of these seasonal periods along with calculated window function shown in top and bottom panel of (b), respectively, with 90\% significance level marked with blue horizontal lines. The highest peak above 90\% significance level indicates the cyclic period of  2.7 $\pm$ 0.1 yr. In bottom panel of (a) each period of 2.7 $\pm$ 0.1 yr is indicated with the vertical lines. The straight lines in each cycle show a linear fit to data during the cycle. The rotation period monotonically decreases along most the star-spot cycles showing a solar-like behaviour. }
\label{fig_diff-rot}
\end{figure*}

%****************************************************************
The visibility of photospheric star-spots is modulated by stellar rotation which causes quasi-periodic brightness variations on time-scales of the order of the rotational period. The modulation period indicates the angular velocity of the latitude at which star-spot activity is predominantly centered. Since the circumpolar spots will not affect the rotational modulation, with an inclination angle ($i$) of 45\deg.0$\pm$2\deg.5 ~on LO Peg \citep{Barnes-05-32, Piluso-08}, any modulation observed on stellar surface would be only due to the spot-groups present within a latitude of $\pm$45\deg ~from the stellar equator. Similar to the solar case, the year to year variations of the rotational period can be described as the migration of stellar activity centres towards latitudes possessing different angular velocity. This migration is caused by the  internal radial shear, which is assumed to be coupled with observed latitudinal shear (e.g. in  $\alpha - \Omega$ dynamo model). 

In order to search for any change in the rotational period, we have determined photometric period of each observing season separately.  We have  chosen the observing  seasons to derive the period due to the fact that the brightness  of star in each season showed regular modulation which could be attributed to rotation of a stationary spot pattern of the star. Smaller time interval and hence smaller baseline introduces large uncertainty in determination of photometric period, whereas larger time intervals shows a significant change in shape of the light curve.  
In case of the sparse data obtained from \hipp ~satellite, the interval were chosen similar to the maximum data length of 0.7 yr obtained from ground based observations. In this way we could obtain 18 seasonal light curves, and average values of each seasonal light curve are shown in the top panel of Fig. \ref{fig_diff-rot} along with the time sequence of $V$-band magnitudes of LO Peg.
 Each seasonal data were analysed using the Scargle-Press period search method. The uncertainty in photometric period and FAP were calculated  following the method of \cite{Horne-86-5}.
In the bottom panel of Fig. \ref{fig_diff-rot}(a), we plot the seasonal values of the measured rotational periods ($P_{\rm sr}$) and the results are summarized in Table \ref{tab_diff-rot}. These modulation periods correspond to the angular velocity of the latitudes at which non-circumpolar spot-groups are present.
Fig. \ref{fig_diff-rot}(b) shows the Scargle power spectra of the measured seasonal rotational periods. Within the Nyquist frequency of 0.00124, we found maximum peak in the Scargle-Press periodogram is above 90\% significance level and has an periodicity of 2.7 $\pm$ 0.1 yr. This period  is well within 3$\sigma$ level of the identified periodicity of brightness variation (see \S ~\ref{subsec:lc-period}). In $\sim$24 yr of observations nine cycles of 2.7 yr period can be made, where we have detected six full cycles and one incomplete cycle (II).  We found that  the rotational period tends to decrease steadily during an `cycle' of $\sim$2.7 yr, and jumping back to a higher value at the beginning of a new cycle.
The abrupt changes in period of cycle-I may be a result of the sparse data set obtained from \hipp ~satellite.  However, in cycle-VII, we did not see any noticeable change in the rotational period.
%*********** fig :6: Optical Multi-band Flare *******************
\begin{figure}
\centering
\includegraphics[width=8.8cm,angle=0]{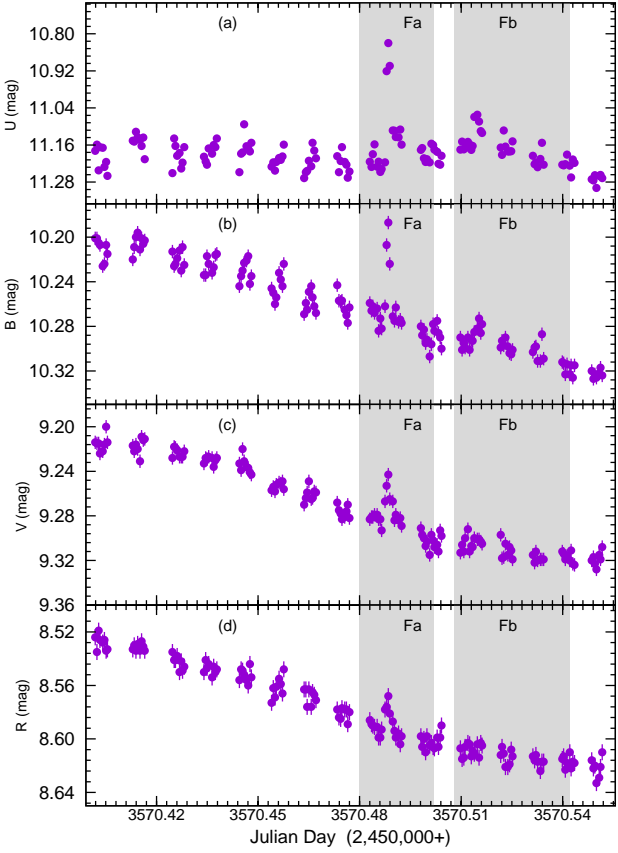}
\caption{Two consecutive representative flares (shaded regions) on LO Peg simultaneously observed in $U$ (a), $B$ (b), $V$ (c) and $R$ (d) optical bands. The first flare shows activity level in all four optical bands. Whereas the second flare is detected only in shorter wavelengths ($U$ and $B$ bands).}
\label{fig_flare-multi}
\end{figure}

%****************************************************************

%%%%%%%%%% SECTION 5 %%%%%%%%%%%%%%%%%%%%%%%%%%%%%%%%%%%%%%%%%%%%%%%%%%%%%%%%%%%%%
\subsection{Flare Analysis}
\label{subsec:op_fl}
 Flares in LO Peg were searched using $U$, $B$, $V$, and $R$ data. For this analysis, we have converted the magnitude into flux using the zero-points given in \cite{Bessel-79-1}. 	     
 Fig. \ref{fig_flare-multi} shows two consecutive representative flares observed simultaneously in all four optical bands. The first flare was detected in all four bands while next flare was not detected in longer wavelengths ($V$ and $R$ band). Thus, a flare detected in one band is not necessarily detected in each of the optical bands.
 Due to the sparse data, we have not followed the usual flare detection methods as described in \cite{Osten-12-3}, \cite{Hawley-14} and \cite{Shibayama-13-2}.  
We have chosen different epochs such that, each epoch contains  a continuous single night observation with at least 16 data points  and  minimum  observing span of $\sim 1$ h. A total of 501 epochs were found using the most populous $V$-band data, among which only 82 epochs have simultaneous observations with other three optical bands.

The light curve of each epoch was first detrended to remove the rotational modulation by fitting a sinusoidal function.
The local mean flux ($F_{\rm lm}$) and standard deviation ($\sigma_{\rm ql}$) of the flux were then computed at each time-sampled data set. 
To avoid misdetection of short stellar brightness enhancement as a flare, candidate flares were flagged as excursions of two or more consecutive data points above 2.5$\sigma_{\rm ql}$ from $F_{\rm lm}$ \citep[see][]{Hawley-14,Davenport-14-1,Lurie-15-1} with at least one of those points being $\geq3\sigma_{\rm ql}$ above $F_{\rm lm}$ in any of the optical band. Once the  flare was detected using the above criteria, the flare segment  was removed to calculate the exact value of  $\sigma_{\rm ql}$, where most of the flares were identified above the 3$\sigma_{\rm ql}$ from the quiescent state. This derived value of $\sigma_{\rm ql}$ was not used for further flare identification.
Finally, each flare candidate, in each photometric band was inspected manually to confirm it as real flares. In this way, we have detected 20 optical flares.  Flare nomenclatures were given as `F$i$', where $i$ = 1, 2, 3, \ldots20; denotes the chronological order of the detected flares.
 Flare parameters of all the detected flares are listed in Table \ref{tab_flare-params}.
%********** tab :3: Optical-Flares ******************************
%\clearpage
%\newgeometry{ left=1cm,right=1cm,top=0.5cm, bottom=0.5cm}
\newcounter{magicrownumbers}
\newcommand\rownumber{\stepcounter{magicrownumbers}\arabic{magicrownumbers}}

\begin{table*}
\scriptsize
\tabcolsep=0.2cm
\begin{adjustwidth}{-0.cm}{}
%\setlength\LTleft{-2cm}
%\begin{center}
\caption{Parameters obtained from flare analysis. }
\label{tab_flare-params} 
\begin{threeparttable}
\begin{tabular}{rcccccccccccr}
\hline \hline \\[-2ex]
   \multicolumn{1}{c}{\textbf{Sl.}} &
   \multicolumn{1}{c}{\textbf{Flare}} &
   \multirow{2}{*}{\textbf{Filter}} &
   \multicolumn{1}{c}{$\mathbf{\mathit{t}_{st}}$$^{a}~$\textbf{(HJD)}} &
   \multicolumn{1}{c}{\textbf{Dn}$^b$} &
   \multicolumn{1}{c}{$\mathbf{\mathit{t}_{pk}}$$^{c}~$\textbf{(HJD)}} &
   \multirow{2}{*}{$\mathbf{\mathit{F}_{lm}}$$^d$} &
   \multirow{2}{*}{$\mathbf{\left(\frac{\mathit{A}_{pk}}{\sigma_{ql}}\right)}$$^e$} &
   \multicolumn{1}{c}{\textbf{\textit{A}}$^f$} &
   \multicolumn{1}{c}{$\mathbf{\mathit{\tau_{r}}}$$^g$} &
   \multicolumn{1}{c}{$\mathbf{\mathit{\tau_{d}}}$$^h$} &
   \multicolumn{1}{c}{\textbf{Energy}$^i$} &
 
\\
%[0.4ex]\cline{4-6}\cline{7-9}\\
   \multicolumn{1}{c}{\textbf{No.}} &
   \multicolumn{1}{c}{\textbf{name}} &
   \multicolumn{1}{c}{} &
   \multicolumn{1}{c}{\textbf{(2400000+)}} &
   \multicolumn{1}{c}{\textbf{(min)}} &
   \multicolumn{1}{c}{\textbf{(2400000+)}} &
   \multicolumn{1}{c}{} &
   \multicolumn{1}{c}{} &
   \multicolumn{1}{c}{\textbf{(frac)}} &
   \multicolumn{1}{c}{\textbf{(min)}} &
   \multicolumn{1}{c}{\textbf{(min)}} &
   \multicolumn{1}{c}{$\mathbf{(10^{32}~erg)}$} &
\\[0.5ex]\hline \hline \\[-2ex]

\rownumber   &  \multirow{2}{*}{F1 }   	 &	 $U$	&	 52857.437	&	26	  &	 52857.445	&	 3.67	&	  3.61	&	 0.133	&	---		          &	10.5     $\pm$	2.6      &    $>$ 2.31	&	 \\  
\rownumber   &                           &	 $B$	&	 52857.437	&	26	  &	 52857.445	&	 11.26	&	  3.37	&	 0.026	&	---		          &	21.9     $\pm$	5.0      &    $>$ 2.81	&	    \\  
           \\                                                                                                                                                                                                                                                     
\rownumber   &  \multirow{3}{*}{F2 }     &	 $U$	&	 53206.468	&	12	  &	 53206.513	&	 3.50	&	  3.03	&	 0.062	&	3.0	  $\pm$	1.2	  &	2.9      $\pm$	1.2      &	 0.57	&	 \\  
\rownumber   &                           &	 $B$	&	 53206.465	&	12	  &	 53206.511	&	 11.11	&	  2.61	&	 0.016	&	$\sim$0.4	          &	$\sim$0.7			 &	 0.09	&	    \\  
           \\                                                                                                                                                                                                                                                     
\rownumber   &  \multirow{2}{*}{F3 }     &	 $U$	&	 53570.317	&	29	  &	 53570.323	&	 3.54	&	  9.50	&	 0.269	&	3.9	  $\pm$	1.4	  &	4.8      $\pm$	1.4      &	 3.25	&	 \\  
\rownumber   &                           &	 $B$	&	 53570.318	&	26	  &	 53570.322	&	 11.03	&	  6.11	&	 0.048	&	1.4	  $\pm$	0.4	  &	1.0      $\pm$	0.4      &	 0.49	&	    \\  
           \\                                                                                                                                                                                                                                                     
\rownumber   &  \multirow{2}{*}{F4 }     &	 $U$	&	 53570.331	&	42	  &	 53570.344	&	 3.45	&	  6.78	&	 0.200	&	6.9	  $\pm$	2.7	  &	$\sim$1.3  	         &	 2.38	&	 \\  
\rownumber   &                           &	 $B$	&	 53570.331	&	42	  &	 53570.341	&	 11.04	&	  5.21	&	 0.040	&	4.9	  $\pm$	1.7	  &	2.0      $\pm$	1.0      &	 1.29	&	    \\  
%\rownumber   &                           &	 V V     &	 53570.330	&	41	  &	 53570.341	&	 25.83	&	  4.44	&	 0.027	&	4.4	  $\pm$	1.8	  &	1.4      $\pm$	0.58     &	 1.73	&	    \\  
           \\                                                                                                                                                                                                                                                     
\rownumber   &  \multirow{4}{*}{F5 }     &	 $U$	&	 53570.359	&	72	  &	 53570.384	&	 3.40	&	  8.42	&	 0.256	&	13.8	  $\pm$	2.2	  &	$\sim$0.7		 &	 5.56	&	\\  
\rownumber   &                           &	 $B$	&	 53570.354	&	72	  &	 53570.376	&	 10.99	&	  4.72	&	 0.036	&	9.8	  $\pm$	2.5	  &	9.2      $\pm$	2.0      &	 3.37	&	    \\  
\rownumber   &                           &	 $V$	&	 53570.354	&	68	  &	 53570.368	&	 25.80	&	  4.48	&	 0.027	&	0.3	  $\pm$	0.1	  &	5.5      $\pm$	1.9      &	 1.80	&	    \\  
\rownumber   &                           &	 $R$	&	 53570.353	&	63	  &	 53570.366	&	 24.46	&	  6.01	&	 0.031	&	2.0	  $\pm$	0.6	  &	7.9      $\pm$	1.3      &	 3.40	&	    \\  
           \\                                                                                                                                                                                                                                                     
\rownumber   &  \multirow{4}{*}{F6 }     &	 $U$	&	 53570.482	&	32	  &	 53570.488	&	 3.22	&	  14.55	&	 0.454	&	1.0	  $\pm$	0.1	  &	2.3      $\pm$	0.3      &	 2.18	&	 \\  
\rownumber   &                           &	 $B$	&	 53570.481	&	23	  &	 53570.488	&	 10.22	&	  10.54	&	 0.090	&	1.3	  $\pm$	0.3	  &	1.2      $\pm$	0.1      &	 0.99	&	    \\  
\rownumber   &                           &	 $V$	&	 53570.482	&	22	  &	 53570.488	&	 23.96	&	  6.52	&	 0.040	&	1.3	  $\pm$	0.3	  &	1.6      $\pm$	0.4      &	 1.26	&	    \\  
\rownumber   &                           &	 $R$	&	 53570.482	&	20	  &	 53570.488	&	 23.00	&	  4.68	&	 0.024	&	1.3	  $\pm$	0.6	  &	1.0      $\pm$	0.4      &	 0.58	&	    \\  
           \\                                                                                                                                                                                                                                                     
\rownumber   &  \multirow{2}{*}{F7 }     &	 $U$	&	 53570.500	&	71	  &	 53570.515	&	 3.17	&	  6.18	&	 0.196	&	8.6	  $\pm$	1.3	  &	18.7     $\pm$	1.6      &	 7.37	&	 \\  
\rownumber   &                           &	 $B$	&	 53570.508	&	49	  &	 53570.515	&	 9.93	&	  3.10	&	 0.027	&	2.3	  $\pm$	0.9	  &	15.0     $\pm$	2.3      &	 2.06	&	    \\  
           \\                                                                                                                                                                                                                                                     
\rownumber   &  \multirow{4}{*}{F8 }     &	 $U$	&	 53971.346	&	52	  &	 53971.359	&	 3.47	&	  10.79	&	 0.302	&	1.7	  $\pm$	0.3	  &	1.2      $\pm$	0.3      &	 1.34	&	 \\  
\rownumber   &                           &	 $B$	&	 53971.339	&	52	  &	 53971.359	&	 10.77	&	  6.10	&	 0.048	&	2.5	  $\pm$	0.9	  &	4.2      $\pm$	0.7      &	 1.55	&	    \\  
\rownumber   &                           &	 $V$	&	 53971.347	&	49	  &	 53971.359	&	 25.17	&	  3.43	&	 0.023	&	2.6	  $\pm$	1.6	  &	5.0      $\pm$	1.4      &	 1.98	&	    \\  
\rownumber   &                           &	 $R$	&	 53971.346	&	49	  &	 53971.358	&	 24.07	&	  2.56	&	 0.015	&	---		  	  &	5.5      $\pm$	2.2      &    $>$ 0.87	&	    \\  
           \\                                                                                                                                                                                                                                                     
\rownumber   &  F9                       &	 $V$	&	 54003.386	&	33	  &	 54003.396	&	 24.48	&	  7.44	&	 0.045	&	4.5	  $\pm$	1.4	  &	5.8      $\pm$	1.6      &	 5.16	&	      \\  
           \\                                                                                                                                                                                                                                                    
\rownumber   &  F10                      &	 $V$	&	 54324.568	&	58	  &	 54324.579	&	 23.98	&	  12.97	&	 0.047	&	---			  &	12.1     $\pm$	1.7      &    $>$ 6.22	&	    \\  
           \\                                                                                                                                                                                                                                                    
\rownumber   &  F11                      &	 $V$	&	 54330.631	&	25	  &	 54330.639	&	 24.58	&	  4.34	&	 0.027	&	---			  &	$\sim$1.0	         &    $>$ 0.32	&	    \\  
           \\                                                                                                                                                                                                                                                    
\rownumber   &  F12                      &	 $V$	&	 54347.557	&	66	  &	 54347.572	&	 24.95	&	  24.25	&	 0.062	&	5.0	  $\pm$	1.0	  &	20.7     $\pm$	3.3      &	 14.63		&	    \\  
           \\                                                                                                                                                                                                                                                    
\rownumber   &  F13                      &	 $V$	&	 54372.403	&      202	  &	 54372.466	&	 24.04	&	  231.62& 	 1.023	&       11.2	  $\pm$	0.7&	22.0     $\pm$	1.3 	 &	 153.61		&	    \\  
           \\                                                                                                                                                                                                                                                     
\rownumber   &  \multirow{4}{*}{F14}     & 	 $U$	&	 54390.293	&	50	  &	 54390.305	&	 3.28	&	  5.56	&	 0.127	&	3.2	  $\pm$	0.7	  &	2.6      $\pm$	0.9      &	 1.10	&	 \\  
\rownumber   &                           & 	 $B$	&	 54390.290	&	50	  &	 54390.307	&	 10.27	&	  8.15	&	 0.041	&	5.8	  $\pm$	0.7	  &	---			 &    $>$ 1.09	&	    \\  
\rownumber   &                           & 	 $V$	&	 54390.293	&	49	  &	 54390.307	&	 24.05	&	  4.19	&	 0.026	&	6.5	  $\pm$	1.4	  &	---		         &    $>$ 1.83	&	    \\  
\rownumber   &                           &	 $R$	&	 54390.294	&	48	  &	 54390.306	&	 23.15	&	  3.28	&	 0.018	&	1.9	  $\pm$	0.9	  &	0.9      $\pm$	0.6      &	 0.54	&	    \\  
           \\                                                                                                                                                                                                                                                     
\rownumber   &  F15                      &	 $U$	&	 54657.392	&	22	  &	 54657.401	&	 3.27	&	  5.36	&	 0.161	&	---			  &	2.3      $\pm$	1.0      &    $>$ 0.54	&	    \\  
           \\                                                                                                                                                                                                                                                     
\rownumber   &  \multirow{2}{*}{F16}     & 	 $U$	&	 54657.503	&	46	  &	 54657.516	&	 3.23	&	  4.15	&	 0.115	&	$\sim$3.2		  &	5.8      $\pm$	1.3      &	 1.48	&	 	\\  
\rownumber   &                           &	 $B$	&	 54657.504	&	43	  &	 54657.517	&	 10.27	&	  3.45	&	 0.028	&	7.3	  $\pm$	2.3	  &	3.6      $\pm$	1.0      &	 1.42	&	    \\  
           \\                                                                                                                                                                                                                                                     
\rownumber   &  F17                      &	 $U$	&	 54747.324	&	33	  &	 54747.336	&	 3.56	&	  3.87	&	 0.067	&	---			  &	1.0      $\pm$	0.4      &    $>$ 0.11	&	    \\  
           \\                                                                                                                                                                                                                                                     
\rownumber   &  \multirow{2}{*}{F18}     & 	 $U$	&	 55070.300	&	26	  &	 55070.307	&	 3.53	&	  4.92	&	 0.097	&	$\sim$2.0	  	  &	2.6      $\pm$	1.3      &	 0.71	&	 \\  
\rownumber   &                           &	 $B$	&	 55070.298	&	26	  &	 55070.308	&	 10.77	&	  3.14	&	 0.017	&	$\sim$2.6	  	  &	$\sim$0.4                & 	 0.26	&	    \\  
           \\                                                                                                                                                                                                                                                    
\rownumber   &  F19                      &	 $U$	&	 55071.261	&	72	  &	 55071.289	&	 3.25	&	  3.04	&	 0.077	&	11.2	  $\pm$	4.5	  &	9.1      $\pm$	2.9      &	 2.26	&	    \\  
           \\                                                                                                                                                                                                                                                     
\rownumber   &  \multirow{4}{*}{F20}     &	 $U$	&	 55071.335	&	57	  &	 55071.350	&	 3.29	&	  3.13	&	 0.078	&	$\sim$4.2		  &	7.8      $\pm$	2.5      &	 1.40	&	 \\  
\rownumber   &                        	 &	 $B$	&	 55071.337	&	56	  &	 55071.349	&	 10.52	&	  5.68	&	 0.040	&	$\sim$0.9		  &	3.7      $\pm$	1.4      &	 0.87	&	    \\  
\rownumber   &                        	 &	 $V$	&	 55071.337	&	55	  &	 55071.350	&	 24.60	&	  3.17	&	 0.022	&	11.1	  $\pm$	2.2	  &	---		         &    $>$ 2.73	&	    \\  
\rownumber   &                        	 &	 $R$	&	 55071.338	&	52	  &	 55071.349	&	 23.73	&	  3.59	&	 0.021	&	9.4	  $\pm$	2.6	  &	---		         &    $>$ 2.09	&	    \\  
           \\    

\hline\hline
\end{tabular}
    \begin{tablenotes}
  \item \textbf{Notes. }
  \item 
$^a$ - Flare start time; $^b$ - Flare duration; $^c$ - Flare peak time; $^d$ - Local mean flux ($F_{lm}$ ) in unit $10^{-11}$ erg s$^{-1}$ cm$^{-2}$; $^e$ - measure of maximum flux increase during flare from quiescent level in multiple of $\sigma_{ql}$; $^f$ - Amplitude of the flare; $^g,^h$ - e-folding rise and decay time of flare; $^i$ - Flare energy. 
 \end{tablenotes}
   \end{threeparttable}
%\end{center}
\normalsize
\end{adjustwidth}
\end{table*}

%****************************************************************
%*********** fig :7: Flare light curves *************************
\begin{figure*}
\centering
% EUO Flares  -->
\subfigure[]{\includegraphics[width=5.1cm,angle=0]{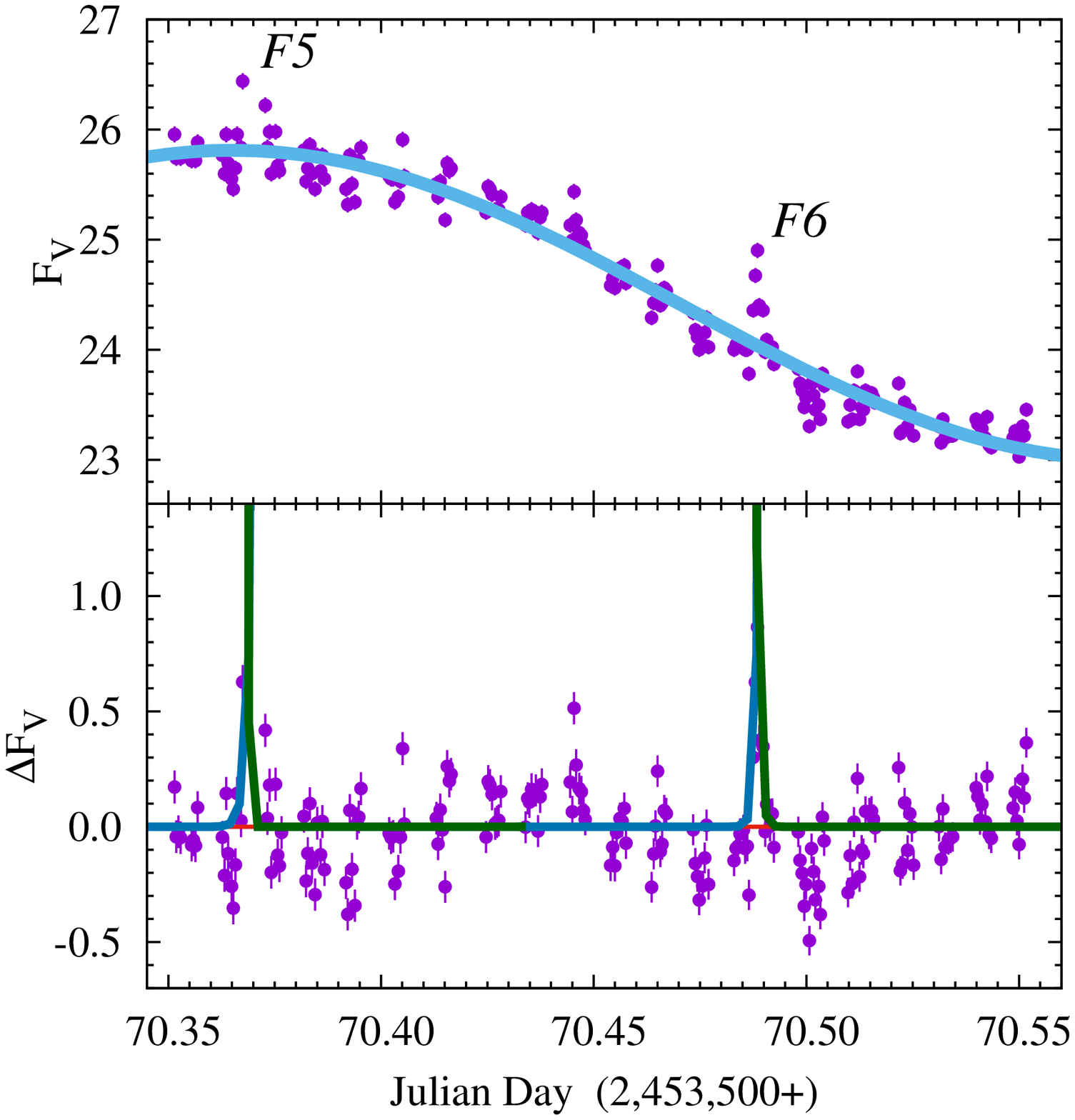}}
\subfigure[]{\includegraphics[width=5.1cm,angle=0]{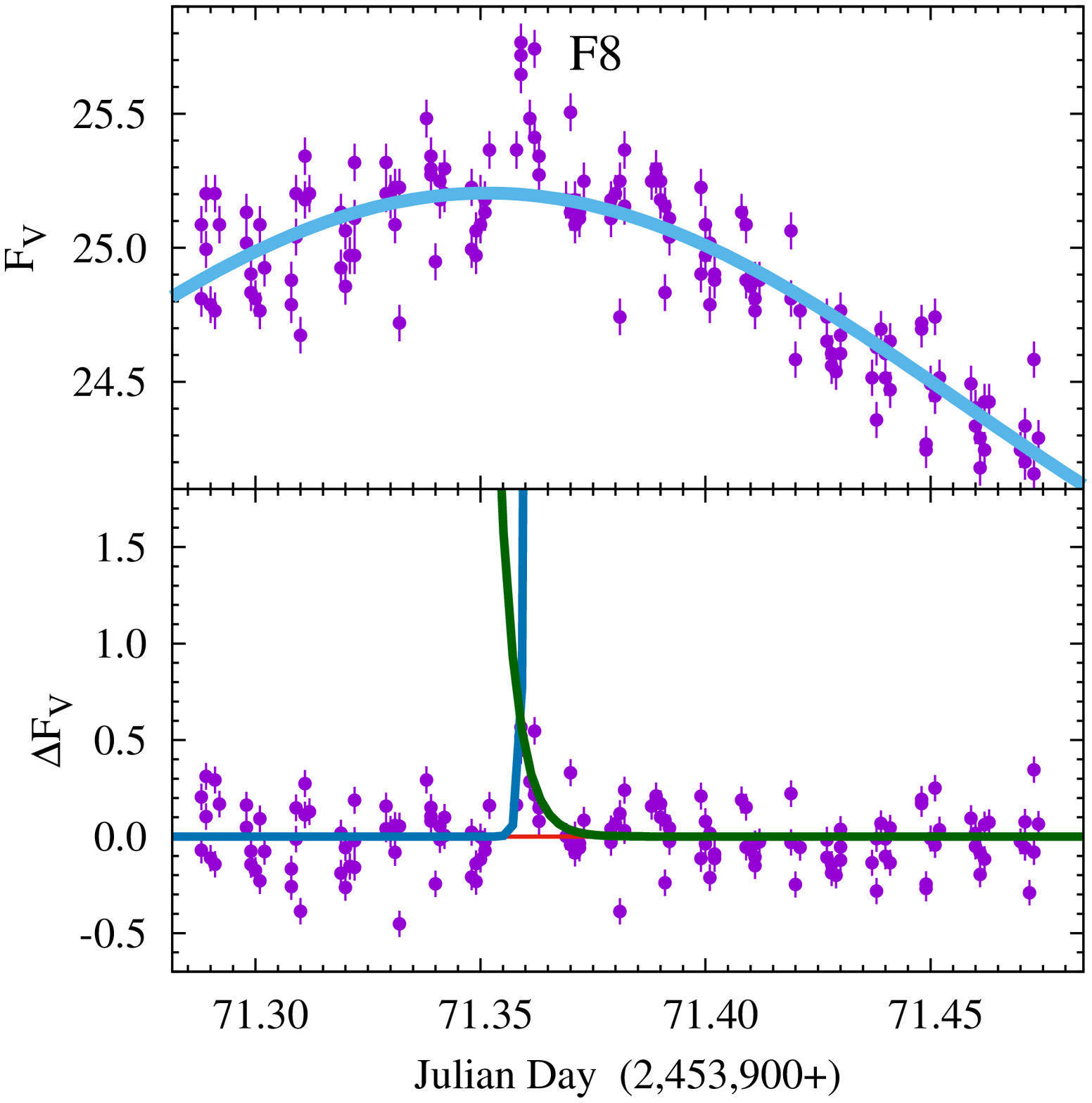}}
% WASP Flares -->
\subfigure[]{\includegraphics[width=5.1cm,angle=0]{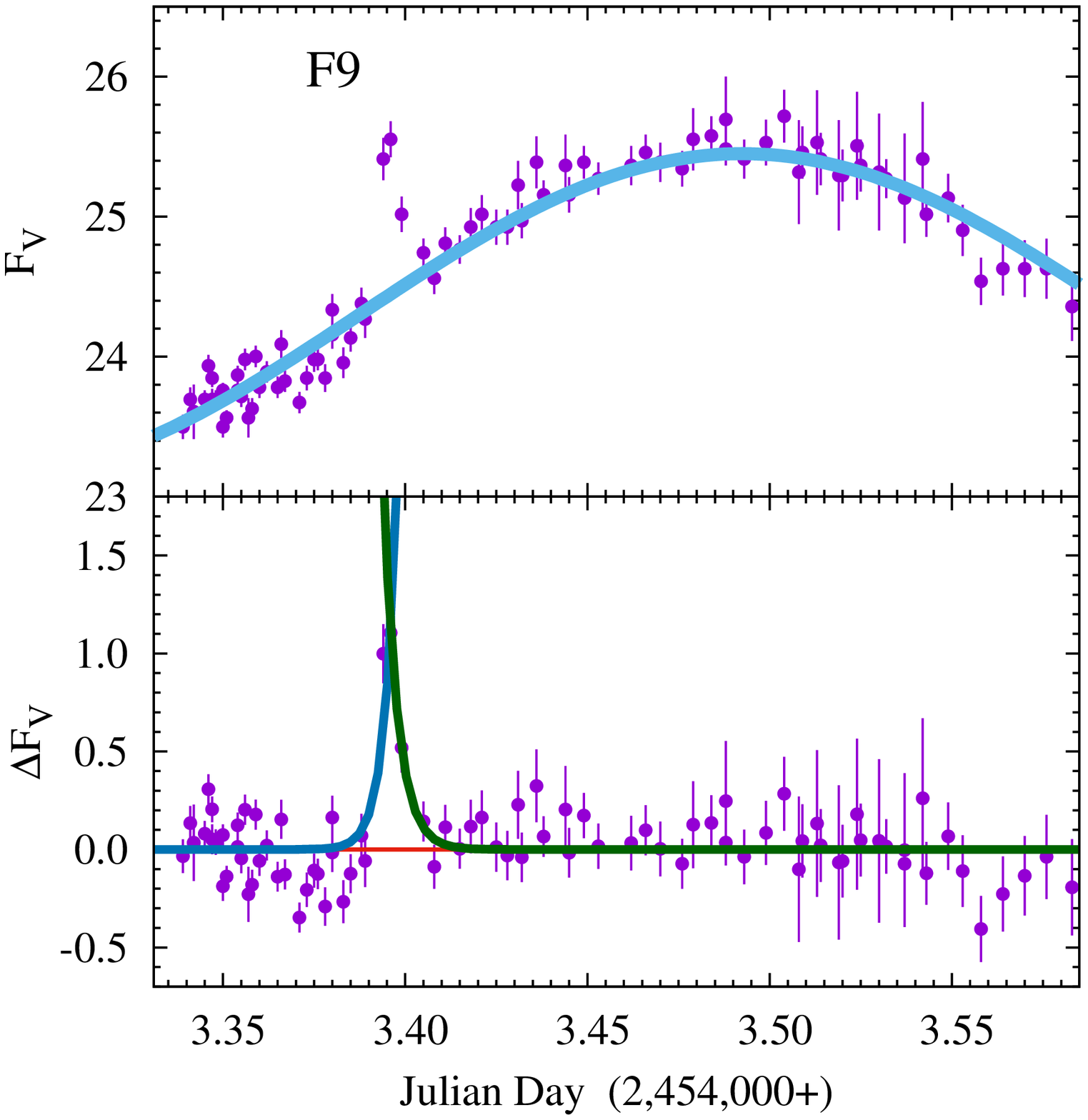}}
\subfigure[]{\includegraphics[width=5.1cm,angle=0]{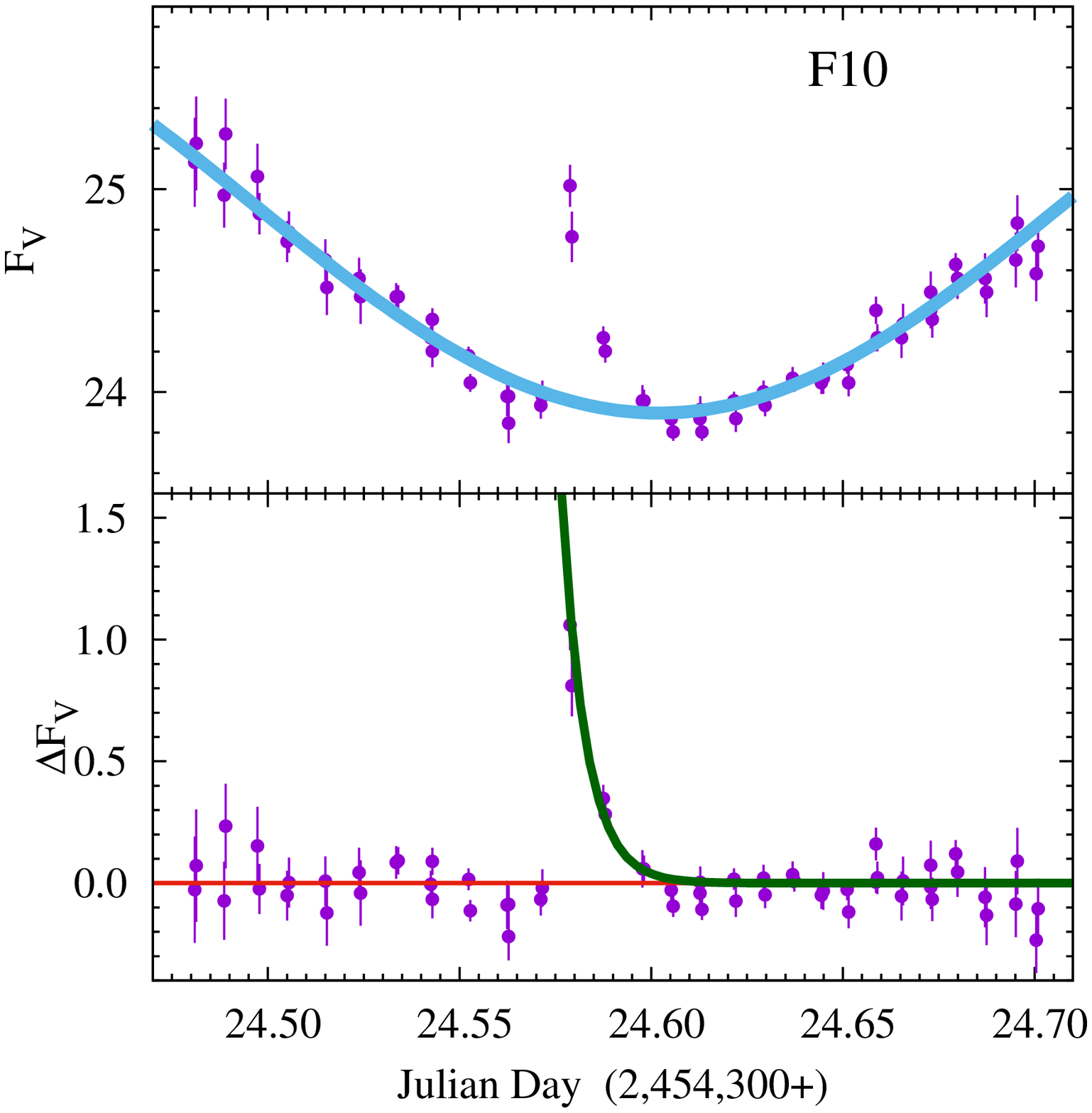}}
\subfigure[]{\includegraphics[width=5.1cm,angle=0]{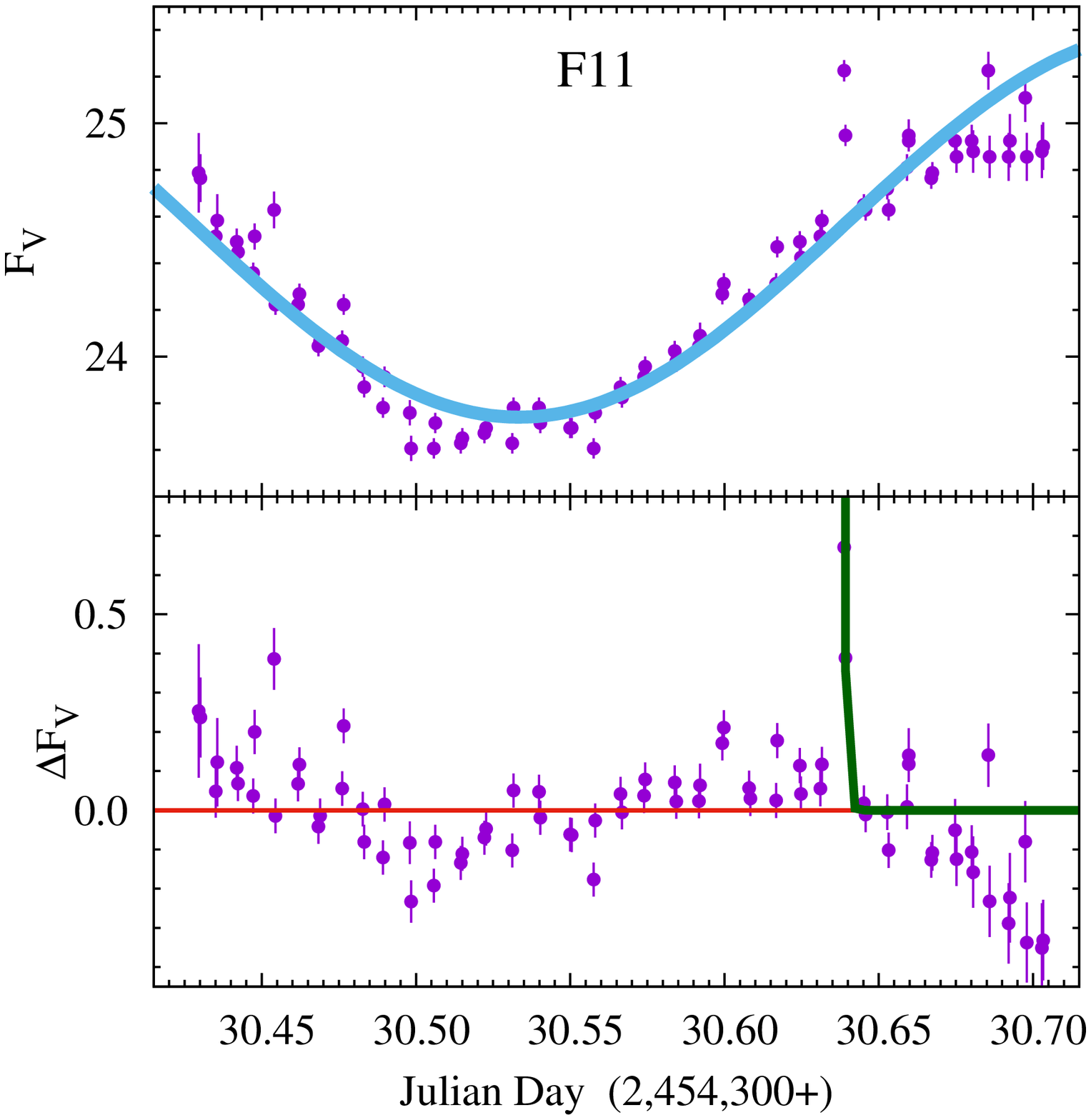}}
\subfigure[]{\includegraphics[width=5.1cm,angle=0]{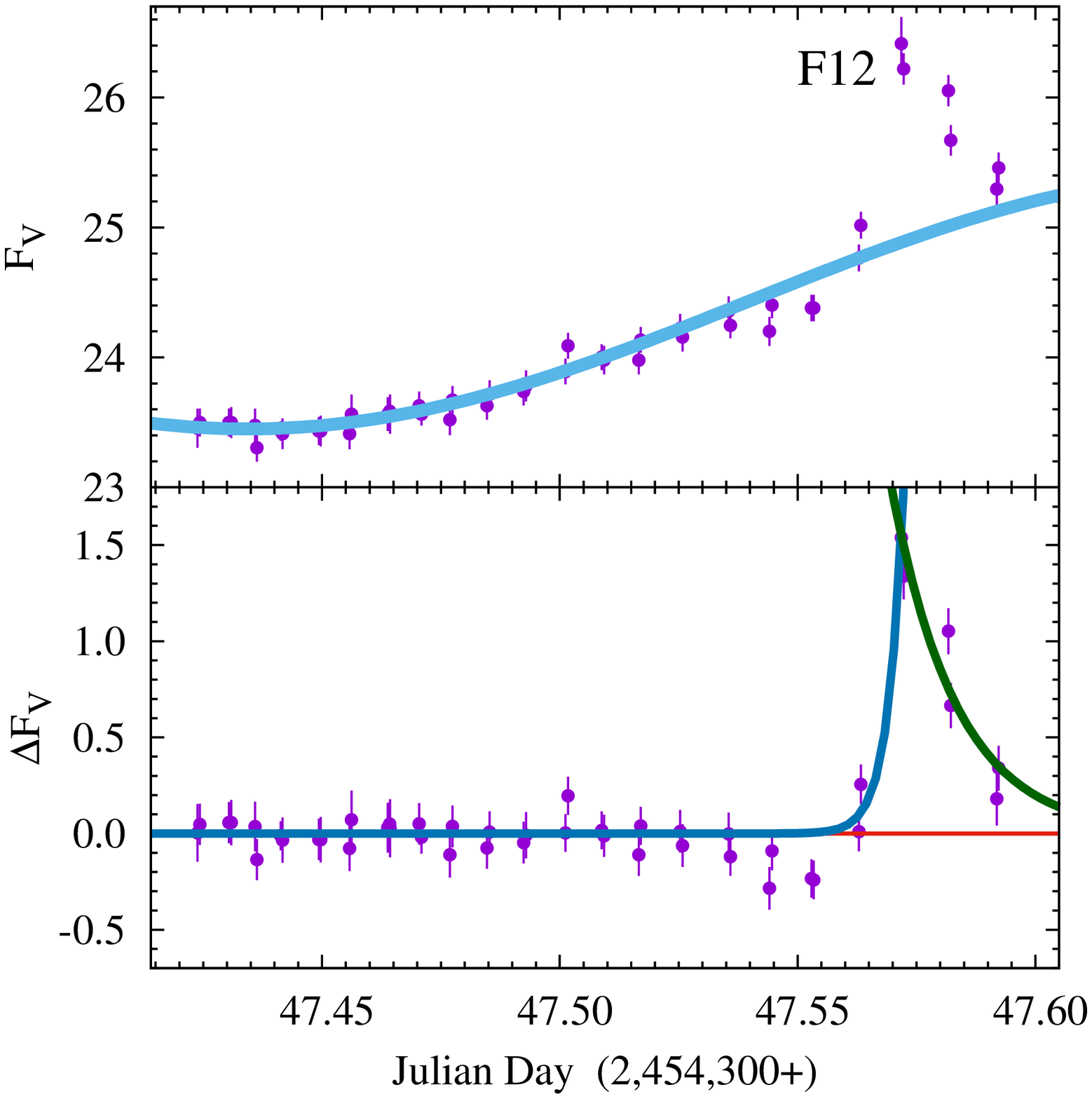}}
\subfigure[]{\includegraphics[width=5.1cm,angle=0]{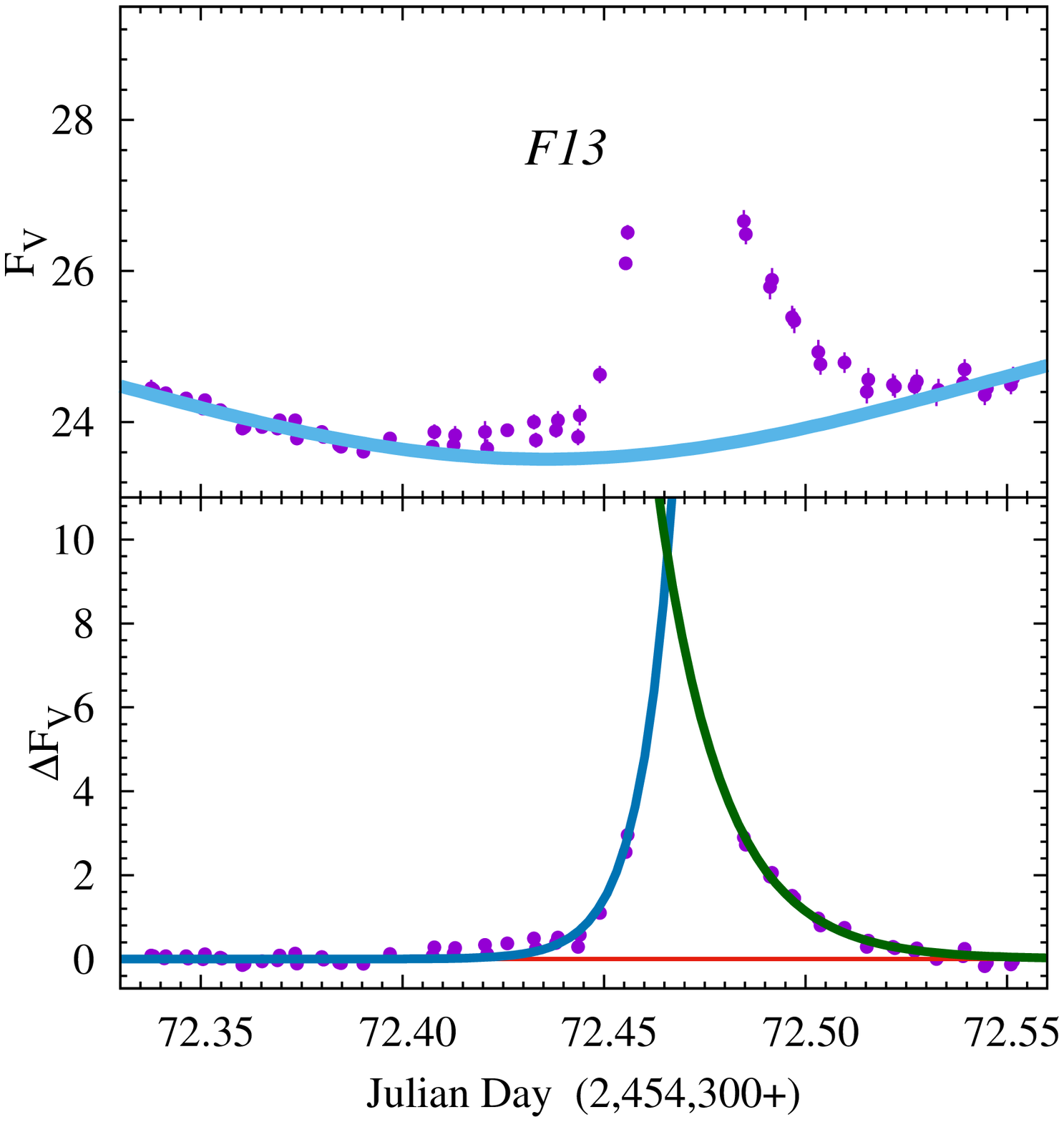}}
% EUO Flares  -->
\subfigure[]{\includegraphics[width=5.1cm,angle=0]{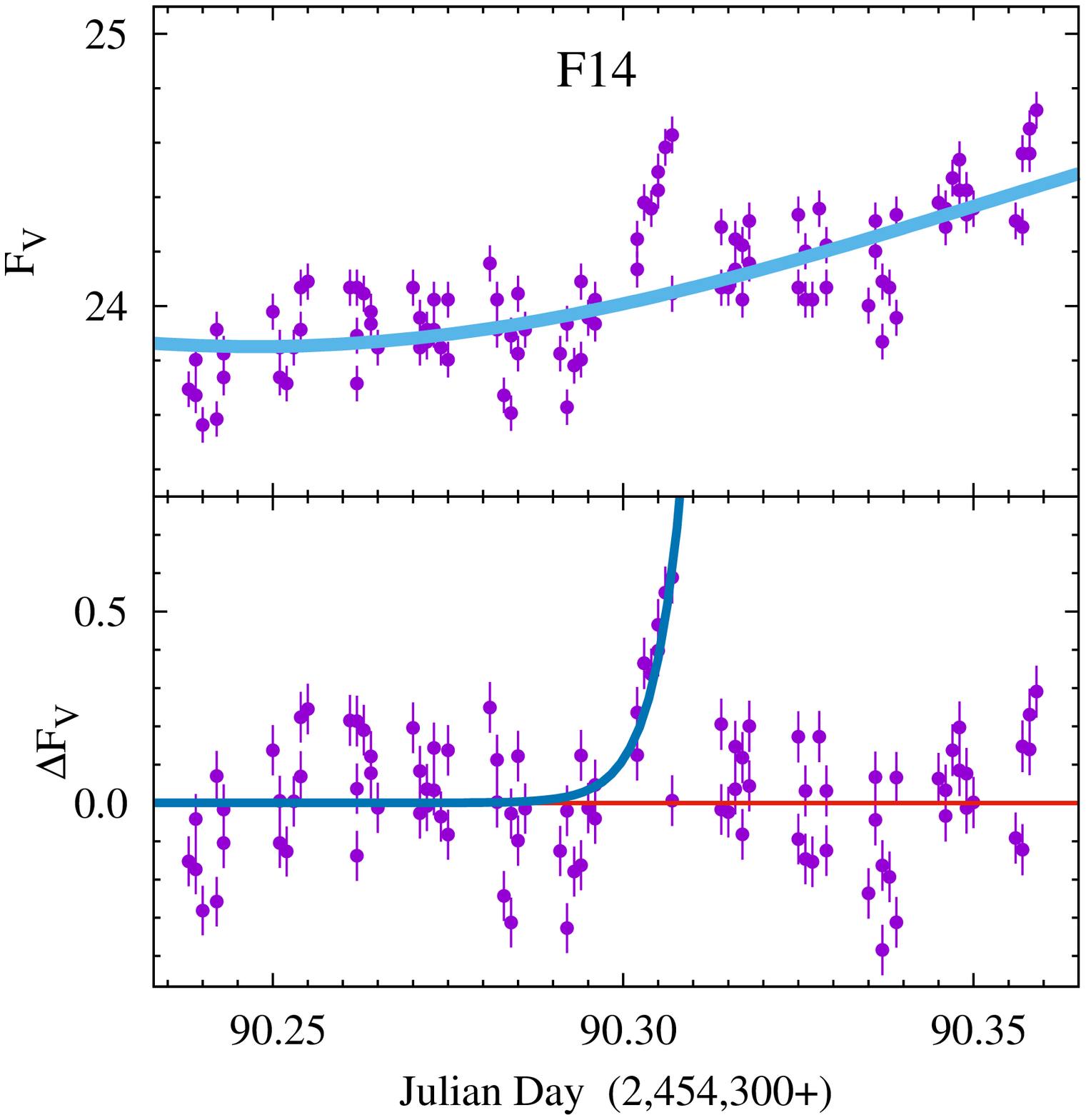}}
\subfigure[]{\includegraphics[width=5.1cm,angle=0]{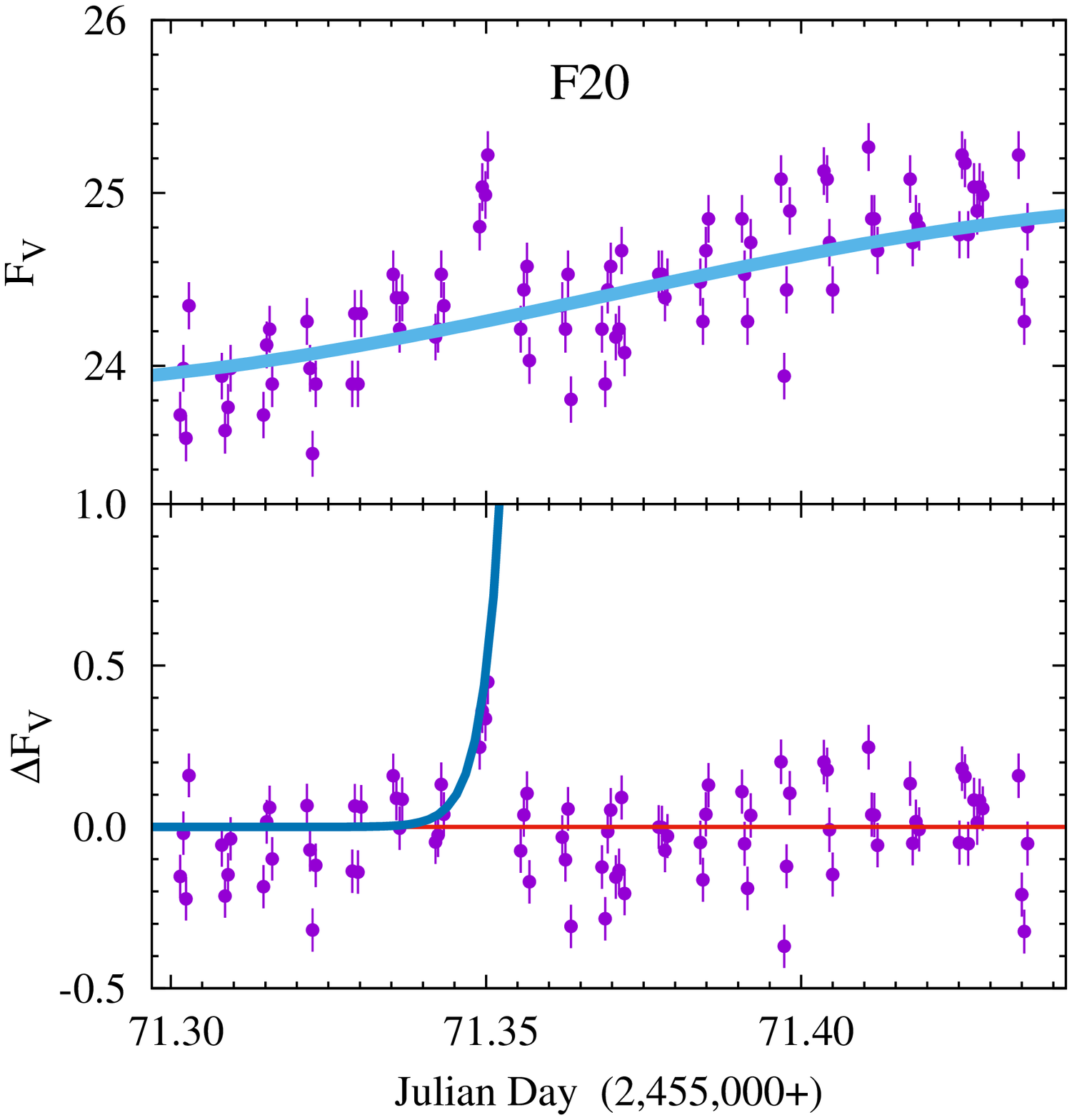}}
\caption{Light curves of all detected $V$-band flares on LO Peg. Top panel of each plot shows the light curve along with best-fitted sinusoid. The bottom panel shows the detrended light curve along with best-fitted exponential functions fitted to flare rise and/or flare decay.}
\label{fig_flare-all}
\end{figure*}

%****************************************************************

 Fig. \ref{fig_flare-all} shows flares detected in $V$-band, where top panels of each plot show $V$-band magnitude variation during flares along with the fitted sinusoidal function %(equation \ref{eqn:sinusoid})
 and bottom panels show the detrended light curve with best fitted exponential function. Most of the flares of LO Peg show usual fast rise (impulsive phase) followed by a slower exponential decay (gradual phase).
 The e-folding rise  (\tr) and decay times (\td) have been derived from the least-squares fit of the exponential function in the form of $F(t) = A_{\rm pk} e^{(t_{\rm pk}-t)/\tau}+F_{\rm lm}$ from flare-start to flare-peak, and from flare-peak to flare-end, respectively. In the fitting procedure  $A_{\rm pk} (= F_{\rm pk} - F_{\rm lm})$, $F_{\rm lm}$ and $t_{\rm pk}$ were fixed parameters. Here, $F_{\rm pk}$ is flux at flare peak at time $t_{\rm pk}$. For the flare F13, the peak was not observed, therefore,  the parameters $A$ and $t_{\rm pk}$ were also kept as free parameters in exponential fitting.
 In order to get a meaningful fit, we restricted our analysis to those flares which contain more than two data points in rise/decay phase. 
 The fitted values of \tr ~and \td ~are given in columns 10 and 11 of Table \ref{tab_flare-params}. The values of \tr ~were found to be in the range of 0.3--14 min with a median value of 2.5 min. Whereas, values of \td~ were derived in the range of 0.4--22 min, with a median of 3.3 min. Most of the time \td~ was found to be more than \tr.

 The amplitude of a flare is defined as 
\begin{equation}
  A  = \frac{A_{\rm pk}}{F_{\rm lm}} = \left( \frac{F_{\rm pk} - F_{\rm lm}}{F_{\rm lm}} \right)
\label{eqn:amplitude}
\end{equation}

 The amplitude of the flare is thus measured relative to the current state of the underlying star, including effects from star-spots, and represents the excess emission above the local mean flux. 
The highest amplitude of 1.02 was found in the long lasting flare F13, while smallest amplitude of 0.016 was found for a small duration flare indicating that long lasting flares are more powerful than small duration flares.
The duration (Dn) of a flare is defined as the difference between the start time (the point in time when the flare flux starts to deviate from the local mean flux) and the end time (when the flare flux returns to the local mean flux). 
The start and end times for each flare were obtained by manual inspection. Flare durations are found within a range of 12--202 min with a median value of 47 min. Flare start time, flare peak time, flare durations, and flare amplitudes are given in fourth, sixth, fifth, and ninth column of Table \ref{tab_flare-params}.
    
The flare energy is computed using the area under the flare light curve i.e. the integrated excess flux ($F_{\rm e}(t)$) released during the flare as
\begin{equation} 
  E_{\rm flare} = 4{\rm \pi} d^{2} \int \! F_{\rm e}(t) \, {\rm d}t
\label{eqn:energy}
\end{equation}
With a distance ($d$) of 25.1 pc for LO Peg \citep{Perryman-97-6}, the derived values of energy in different filters for all detected flares are given in column 12 of Table \ref{tab_flare-params}. The flare energies are found in between 9 \E{30} erg and 1.54 \E{34} erg. The most energetic flare is the longest flare.
Since the total energy released by the flare must be smaller than (or equal to) the magnetic energy stored around the star-spots (i.e. $E_{\rm flare}\leq E_{\rm mag}$), the minimum magnetic field can be estimated during the flare as $E_{\rm mag}~\alpha~B^{2}l^{3}$. Assuming the loop-length of typical flares on G-K stars are of the order of \Pten{10} cm \citep[see][]{Gudel-01-4, Pandey-08-4}. The minimum magnetic field in the observed flares are estimated to be  0.1--3.5 kG. 

%%%%%%%%%% SECTION 6 %%%%%%%%%%%%%%%%%%%%%%%%%%%%%%%%%%%%%%%%%%%%%%%%%%%%%%%%%%%%%
\subsection{Surface imaging with light curve inversion technique}
\label{subsec:si_li}
%*********** fig :8: Surface Imaging ****************************
\begin{figure*}
\centering
\includegraphics[width=17.1cm,angle=0,trim={0cm 0cm 0cm 1cm},clip=false]{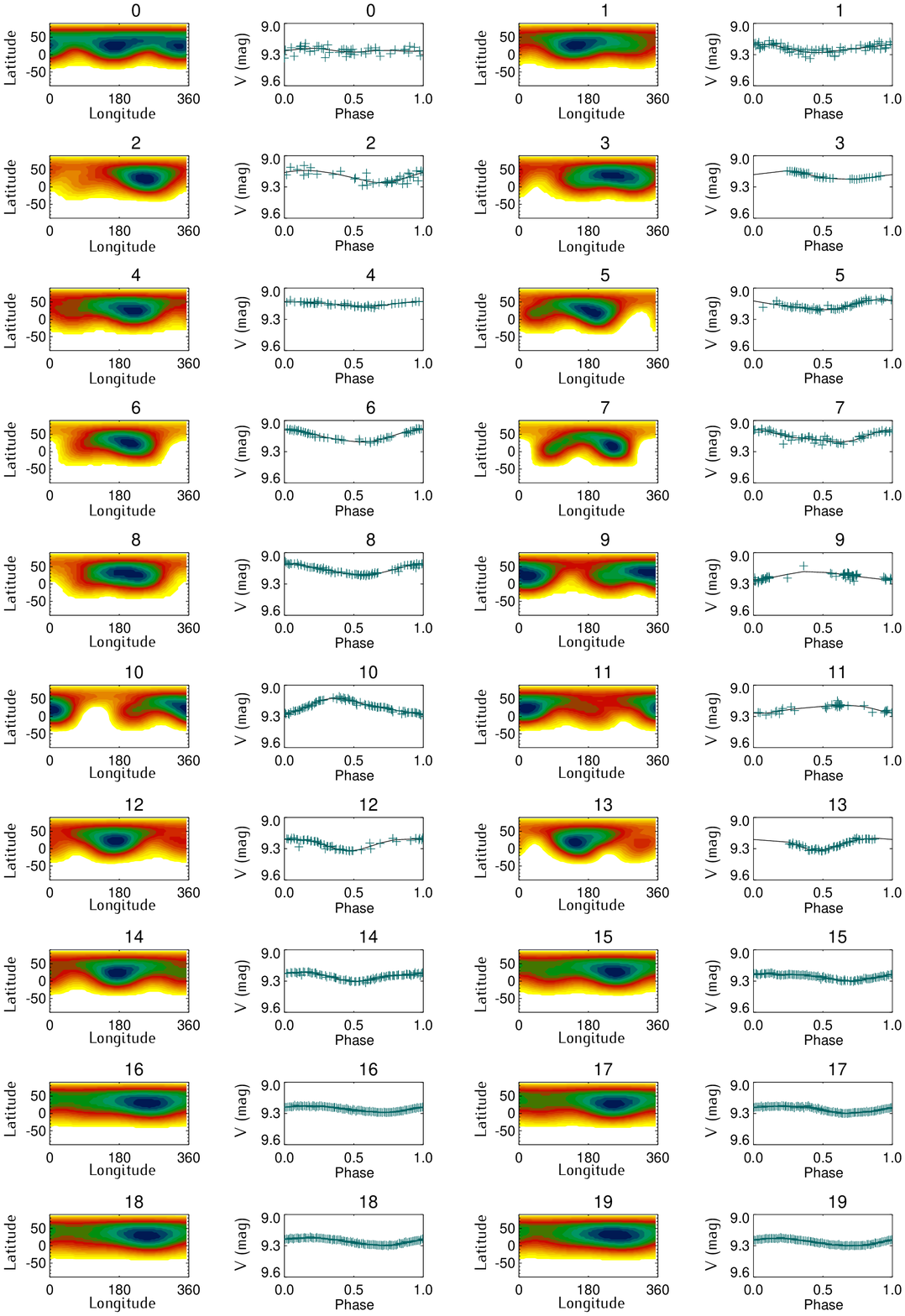}
\caption{The surface temperature inhomogeneity maps of LO Peg --- Continued} 
\end{figure*}
%------------ Next page-----------------------
\addtocounter{figure}{-1}
\begin{figure*}
\centering
\includegraphics[width=17.1cm,angle=0,trim={0cm 0cm 0cm 1cm},clip=false]{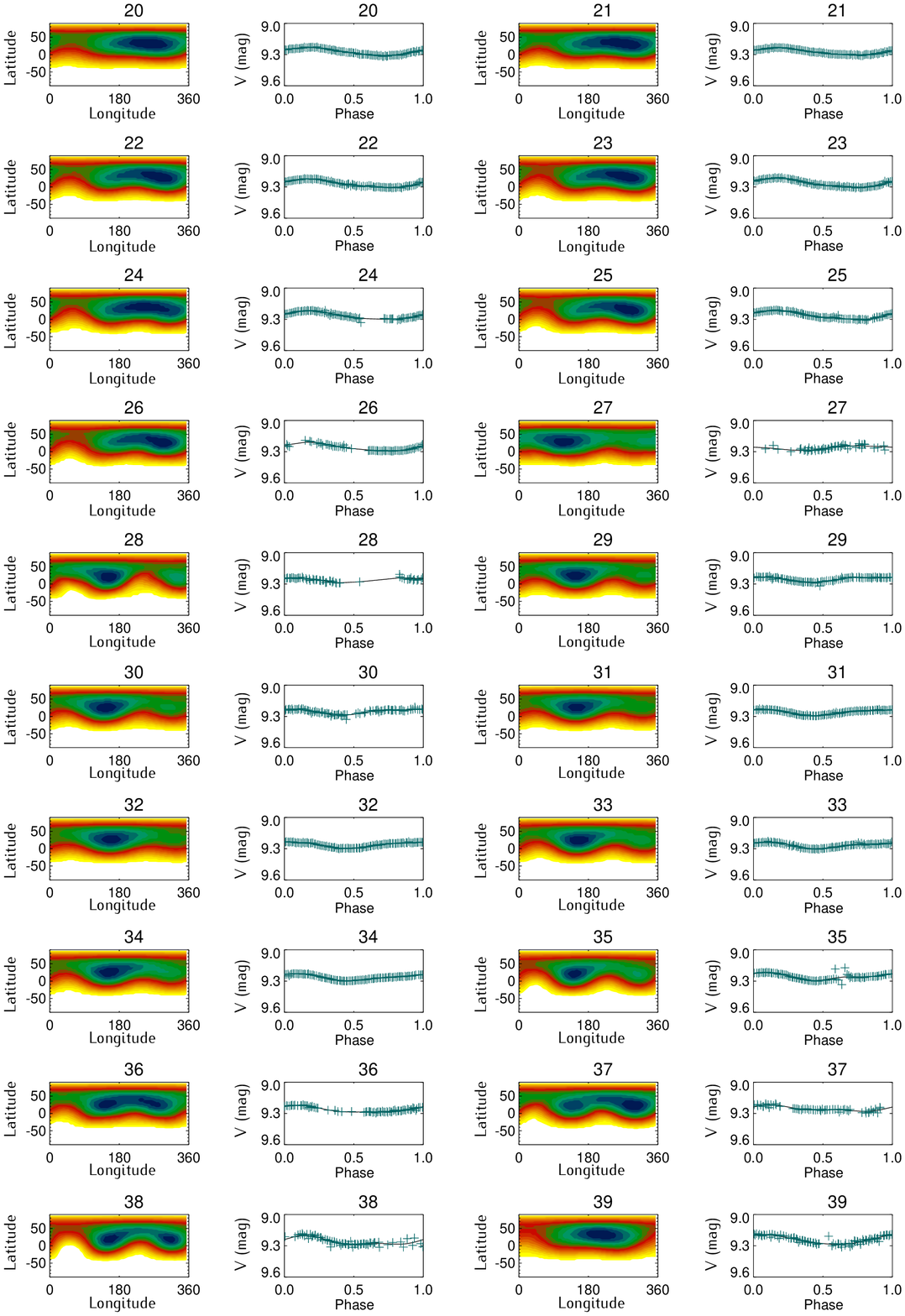}
\caption{The surface temperature inhomogeneity maps of LO Peg --- Continued} 
\end{figure*}
%------------ Next page-----------------------
\addtocounter{figure}{-1}
\begin{figure*}
\centering
\includegraphics[width=17.1cm,angle=0,trim={0cm 0cm 0cm 18cm},clip=false]{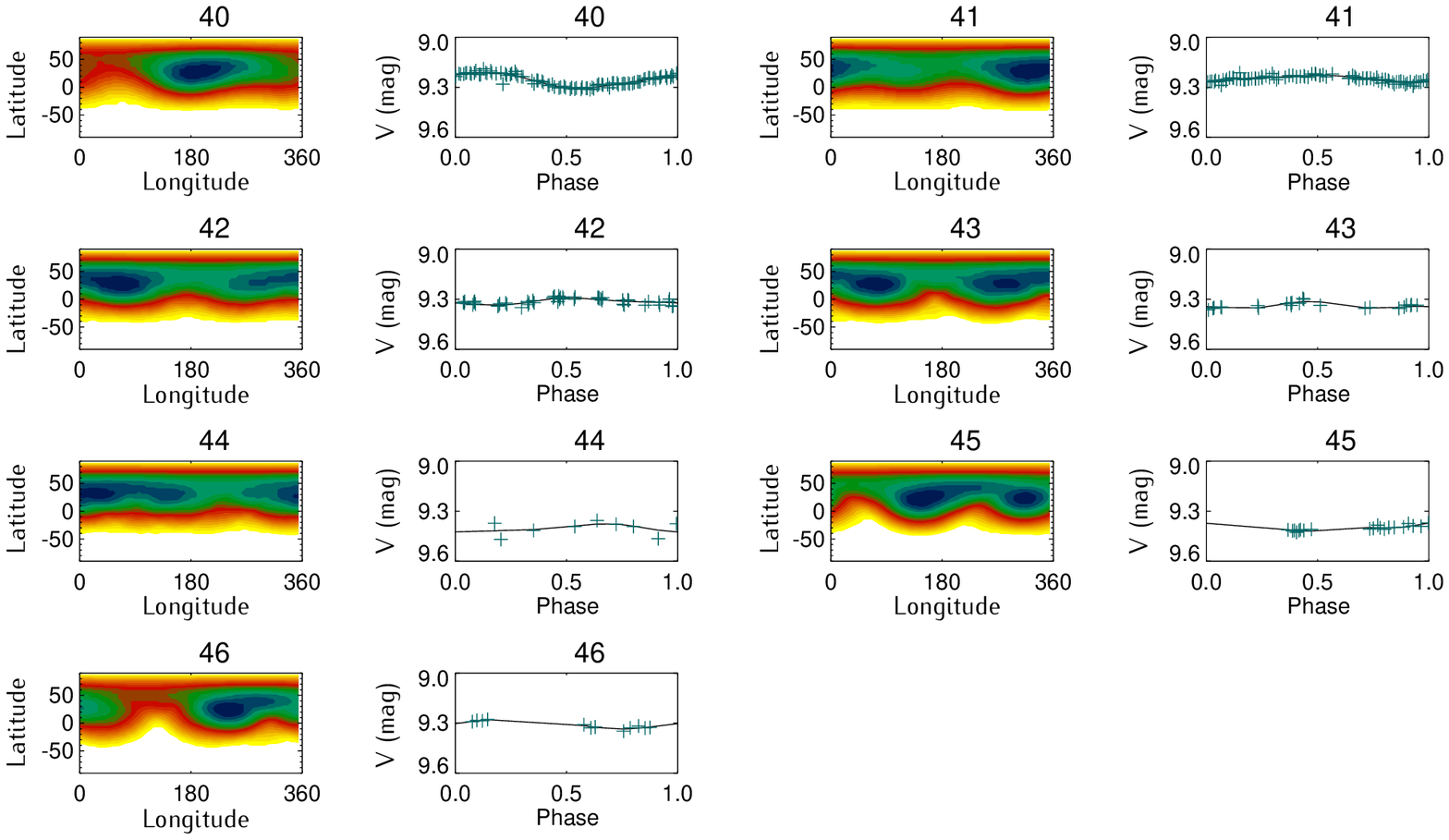}
\caption{The surface temperature inhomogeneity maps of LO Peg for 47 epochs are shown in left-hand panels (first and third columns). The surface maps are presented on the same scale, with darker regions corresponding to higher spot-filling factors. Right-hand panels (second and fourth columns) show the light curves folded on each epoch. Observed and calculated $V$-band light curves are presented by crosses and continuous lines, respectively.} 
\label{fig_surface-plot}
\end{figure*}

%****************************************************************
In order to determine locations of spots on the stellar surface, we have performed inversion of the phased light curves into stellar images using the light curve inversion code \citep[{\sc iph}; see][]{Savanov-08-2, Savanov-11-4}. The model assumes that, due to the low spatial resolution, the local intensity of the stellar surface always has a contribution from the photosphere ($I_{\rm P}$) and from cool spots ($I_{\rm S}$) weighted by the fraction of the surface covered by spots, i.e., the spot filling factor $f$ by the following relation:  $I = f \times I_{\rm P} + (1 - f ) \times I_{\rm S}$ ; with $0 < f < 1$. The inversion of a light curve results in a distribution of the spot-filling factor ($f$) over the visible stellar surface. Although this approach is less informative than the Doppler imaging technique \citep[see][]{Strassmeier-00-7,  Barnes-05-32, Piluso-08}; however, analysis of long time series of photometric observations allows us to recover longitudinal spot patterns and study of their long-term evolution. 

We could make 47 time intervals by manual inspection such that each interval had a sufficient number of data points and had no noticeable changes in their shape.
Individual light curves were analysed using the {\sc iph} code. Several sets of time interval contain a large number of observations within it (e.g. set 33 includes 1007 measurement), in those cases we divided the time axis of the phase diagram into 100 bins and averaged the measurements. In our modelling, the surface of the star was divided into a grid of $6^{\circ}\times 6^{\circ}$ pixels (unit areas), and the values of $f$ were determined for each grid pixel. We adopt the photospheric temperature of LO Peg to be $\sim$4500 K \citep[see][]{Pandey-05-8} and the spot temperature to be 750 K lower than the photospheric temperature \citep[][]{Piluso-08, Savanov-11-4}. The stellar astrophysical input includes a set of photometric fluxes calculated from atmospheric model by \cite{Kurucz-92-5} as a function of temperature and gravity.
For LO Peg the `$i$' was precisely determined with the analysis of a very extensive set of high resolution spectra \cite[see][]{Barnes-05-32,Piluso-08}, therefore, in our reconstruction of the temperature inhomogeneity maps we safely fixed the inclination angle at 45\deg.
Various test cases were performed to recover the artificial maps and include data errors and different input parameter errors which demonstrate the robustness of our solution to various false parameters. 

%********** tab :4: Surface imaging parameters ******************
\begin{table*}
\begin{center}
\scriptsize
\tabcolsep=0.25cm
\caption{ Parameters derived from light curve modelling.}
\label{tab_surface-params}
\begin{tabular}{ccccccccccccc}
\hline\hline
{\bf Epoch} &	 {\bf No. of } 	& {\bf HJD$_{\rm beg}$}   & {\bf HJD$_{\rm end}$}  & {\bf HJD$_{\rm middle}$}& {V$_{\rm max}$} 	&   {V$_{\rm min}$}	& {V$_{\rm mean}$}	&         $A_{\rm LI}$	&   {\bf Sp} &    $\psi_{1}$	&    $\psi_{2}$	&	Note$^\dagger$ \\  
{\bf } 	&	 {\bf Points} 	& {\bf (2400000+) }   & {\bf (2400000+) }  & {\bf (2400000+) }  &	 (mag)  &        (mag)  &       (mag)  &       (mag)  &     (\%) 	&  ($^{\circ}$) &  ($^{\circ}$) &	         \\  
\hline\hline
0 	&	 47 	&	 47857.50 	&	 48458.42 	&	 48157.96 	&	 9.193 	&	 9.333 	&	 9.260 	&	 0.140 	&	 16.6	&	 172 	&	 342 	&	 e\\  
1 	&	 74	&	 48550.46	&	 48717.97	&	 48634.22	&	 9.167	&	 9.337	&	 9.237	&	 0.170	&	 15.2	&	 148	&	 285	&	 n\\  
2 	&	 39	&	 48788.25	&	 48972.28	&	 48880.26	&	 9.094	&	 9.287	&	 9.204	&	 0.193	&	 11.9	&	 252	&	 --	&	 n\\  
3 	&	 37	&	 52181.17	&	 52198.13	&	 52189.65	&	 9.147	&	 9.228	&	 9.191	&	 0.081	&	 11.2	&	 248	&	 --	&	 u\\  
4 	&	 50	&	 52546.21	&	 52551.25	&	 52548.73	&	 9.116	&	 9.192	&	 9.154	&	 0.076	&	  8.8	&	 222	&	 --	&	 n\\  
5 	&	 232	&	 52755.91	&	 52860.68	&	 52808.29	&	 9.091	&	 9.221	&	 9.165	&	 0.130	&	  9.6	&	 198	&	 45	&	 n\\  
6 	&	 204	&	 52863.39	&	 52872.41	&	 52867.90	&	 9.071	&	 9.217	&	 9.135	&	 0.146	&	  8.8	&	 212	&	 --	&	 n\\  
7 	&	 168	&	 52874.64	&	 52890.55	&	 52882.59	&	 9.067	&	 9.235	&	 9.141	&	 0.168	&	  9.3	&	 248	&	 88	&	 n\\  
8 	&	 290	&	 52893.59	&	 52942.54	&	 52918.07	&	 9.075	&	 9.254	&	 9.160	&	 0.179	&	  9.4	&	 210	&	 --	&	 n\\  
9 	&	 74	&	 53142.92	&	 53199.49	&	 53171.21	&	 9.126	&	 9.288	&	 9.233	&	 0.162	&	 13.3	&	 20	&	 260	&	 u\\  
10	&	 464	&	 53203.42	&	 53211.47	&	 53207.44	&	 9.098	&	 9.297	&	 9.205	&	 0.199	&	 12.7	&	 3	&	 210	&	 n\\  
11	&	 36	&	 53212.39	&	 53344.52	&	 53278.46	&	 9.150	&	 9.287	&	 9.225	&	 0.137	&	 14.1	&	 18	&	 190	&	 n\\  
12	&	 221	&	 53487.92	&	 53570.69	&	 53529.30	&	 9.176	&	 9.353	&	 9.252	&	 0.177	&	 15.8	&	 173	&	 --	&	 n\\  
13	&	 242	&	 53571.30	&	 53574.69	&	 53572.99	&	 9.168	&	 9.333	&	 9.259	&	 0.165	&	 14.5	&	 148	&	 330	&	 u\\  
14	&	 426	&	 53584.66	&	 53650.35	&	 53617.51	&	 9.197	&	 9.316	&	 9.253	&	 0.119	&	 16.1	&	 178	&	 --	&	 n\\  
15	&	 746	&	 53853.92	&	 53938.68	&	 53896.30	&	 9.207	&	 9.331	&	 9.255	&	 0.124	&	 16.4	&	 258	&	 80	&	 n\\  
16	&	 555	&	 53943.57	&	 53955.73	&	 53949.65	&	 9.200	&	 9.302	&	 9.257	&	 0.102	&	 16.3	&	 260	&	 --	&	 n\\  
17	&	 315	&	 53960.41	&	 53963.44	&	 53961.93	&	 9.216	&	 9.307	&	 9.264	&	 0.091	&	 16.5	&	 250	&	 --	&	 n\\  
18	&	 308	&	 53966.46	&	 53969.69	&	 53968.08	&	 9.210	&	 9.303	&	 9.258	&	 0.093	&	 16.3	&	 253	&	 --	&	 n\\  
19	&	 638	&	 53970.41	&	 53973.68	&	 53972.04	&	 9.206	&	 9.315	&	 9.256	&	 0.109	&	 16.8	&	 252	&	 --	&	 n\\  
20	&	 599	&	 53975.40	&	 53981.55	&	 53978.48	&	 9.194	&	 9.328	&	 9.276	&	 0.134	&	 17.4	&	 260	&	 --	&	 n\\  
21	&	 504	&	 53987.40	&	 53994.62	&	 53991.01	&	 9.217	&	 9.316	&	 9.270	&	 0.099	&	 17.4	&	 275	&	 --	&	 n\\  
22	&	 241	&	 53995.36	&	 53998.61	&	 53996.98	&	 9.212	&	 9.310	&	 9.258	&	 0.098	&	 17.1	&	 295	&	 180	&	 n\\  
23	&	 286	&	 54001.34	&	 54005.59	&	 54003.47	&	 9.204	&	 9.320	&	 9.264	&	 0.116	&	 16.8	&	 280	&	 180	&	 n\\  
24	&	 228	&	 54006.39	&	 54011.57	&	 54008.98	&	 9.206	&	 9.331	&	 9.254	&	 0.125	&	 16.9	&	 255	&	 --	&	 n\\  
25	&	 242	&	 54017.33	&	 54023.38	&	 54020.36	&	 9.202	&	 9.319	&	 9.257	&	 0.117	&	 16.7	&	 290	&	 175	&	 n\\  
26	&	 130	&	 54029.40	&	 54044.46	&	 54036.93	&	 9.191	&	 9.300	&	 9.274	&	 0.109	&	 16.3	&	 300	&	 190	&	 n\\  
27	&	 181	&	 54227.90	&	 54281.78	&	 54254.84	&	 9.218	&	 9.311	&	 9.266	&	 0.093	&	 16.8	&	 115	&	 345	&	 n\\  
28	&	 243	&	 54283.77	&	 54289.74	&	 54286.75	&	 9.207	&	 9.303	&	 9.256	&	 0.096	&	 16.1	&	 155	&	 335	&	 u\\  
29	&	 266	&	 54293.75	&	 54305.73	&	 54299.74	&	 9.223	&	 9.319	&	 9.254	&	 0.096	&	 16.0	&	 150	&	 330	&	 n\\  
30	&	 428	&	 54306.49	&	 54312.69	&	 54309.59	&	 9.204	&	 9.330	&	 9.251	&	 0.126	&	 16.1	&	 150	&	 315	&	 n\\  
31	&	 686	&	 54315.69	&	 54328.70	&	 54322.19	&	 9.217	&	 9.309	&	 9.261	&	 0.092	&	 16.5	&	 152	&	 315	&	 n\\  
32	&	 1007	&	 54329.43	&	 54340.68	&	 54335.06	&	 9.206	&	 9.331	&	 9.261	&	 0.125	&	 16.7	&	 160	&	 310	&	 n\\  
33	&	 557	&	 54343.59	&	 54352.64	&	 54348.12	&	 9.179	&	 9.327	&	 9.267	&	 0.148	&	 17.0	&	 155	&	 320	&	 n\\  
34	&	 546	&	 54353.37	&	 54364.53	&	 54358.95	&	 9.219	&	 9.307	&	 9.268	&	 0.088	&	 16.8	&	 158	&	 300	&	 n\\  
35	&	 320	&	 54368.37	&	 54377.43	&	 54372.90	&	 9.169	&	 9.334	&	 9.258	&	 0.165	&	 16.3	&	 148	&	 312	&	 n\\  
36	&	 333	&	 54381.44	&	 54394.27	&	 54387.86	&	 9.203	&	 9.303	&	 9.265	&	 0.100	&	 16.8	&	 150	&	 275	&	 e\\  
37	&	 133	&	 54405.31	&	 54454.05	&	 54429.68	&	 9.196	&	 9.306	&	 9.252	&	 0.110	&	 16.2	&	 145	&	 308	&	 n\\  
38	&	 224	&	 54590.92	&	 54661.78	&	 54626.35	&	 9.174	&	 9.338	&	 9.254	&	 0.164	&	 16.4	&	 158	&	 318	&	 n\\  
39	&	 385	&	 54663.77	&	 54710.68	&	 54687.22	&	 9.169	&	 9.332	&	 9.234	&	 0.163	&	 15.3	&	 218	&	 --	&	 n\\  
40	&	 452	&	 54716.66	&	 54785.52	&	 54751.09	&	 9.189	&	 9.339	&	 9.260	&	 0.150	&	 16.4	&	 195	&	 --	&	 n\\  
41	&	 339	&	 54954.92	&	 55105.62	&	 55030.27	&	 9.191	&	 9.307	&	 9.253	&	 0.116	&	 15.7	&	 120	&	 328	&	 n\\  
42	&	 47	&	 55130.10	&	 55196.05	&	 55163.07	&	 9.281	&	 9.350	&	 9.313	&	 0.069	&	 20.1	&	 65	&	 295	&	 n\\  
43	&	 23	&	 55489.16	&	 55526.10	&	 55507.63	&	 9.295	&	 9.364	&	 9.338	&	 0.069	&	 21.8	&	 73	&	 292	&	 u\\  
44	&	 9	&	 55758.82	&	 55775.77	&	 55767.29	&	 9.356	&	 9.469	&	 9.402	&	 0.113	&	 25.7	&	 25	&	 145	&	 u\\  
45	&	 20	&	 56239.14	&	 56257.17	&	 56248.16	&	 9.370	&	 9.427	&	 9.402	&	 0.057	&	 24.3	&	 153	&	 322	&	 u\\  
46	&	 12	&	 56636.36	&	 56645.36	&	 56640.86	&	 9.278	&	 9.348	&	 9.312	&	 0.070	&	 18.6	&	 248	&	 5	&	 u\\  
 
\hline \hline   
\multicolumn{13}{l}{{\bf Notes.}}\\
\multicolumn{13}{l}{HJD$_{\rm beg}$, HJD$_{\rm end}$, and HJD$_{\rm middle}$ are start, end, and middle time of each epoch. V$_{\rm max}$, V$_{\rm min}$, and V$_{\rm mean}$ are maximum, minimum and mean }\\
\multicolumn{13}{l}{magnitudes of LO Peg. $A_{\rm LI}$ is the amplitude of variability. Sp is the spottedness of the stellar surface. $\psi_{1}$ and $\psi_{2}$ are active longitudes.}\\
\multicolumn{13}{l}{$^\dagger$ e -- size of both spots were approximately equal;  n -- size of both spots were different; u -- uncertain results due to incomplete light curve.}
\end{tabular}
\end{center}
\end{table*}

\begin{figure*}
\includegraphics[width=18.5cm,angle=0]{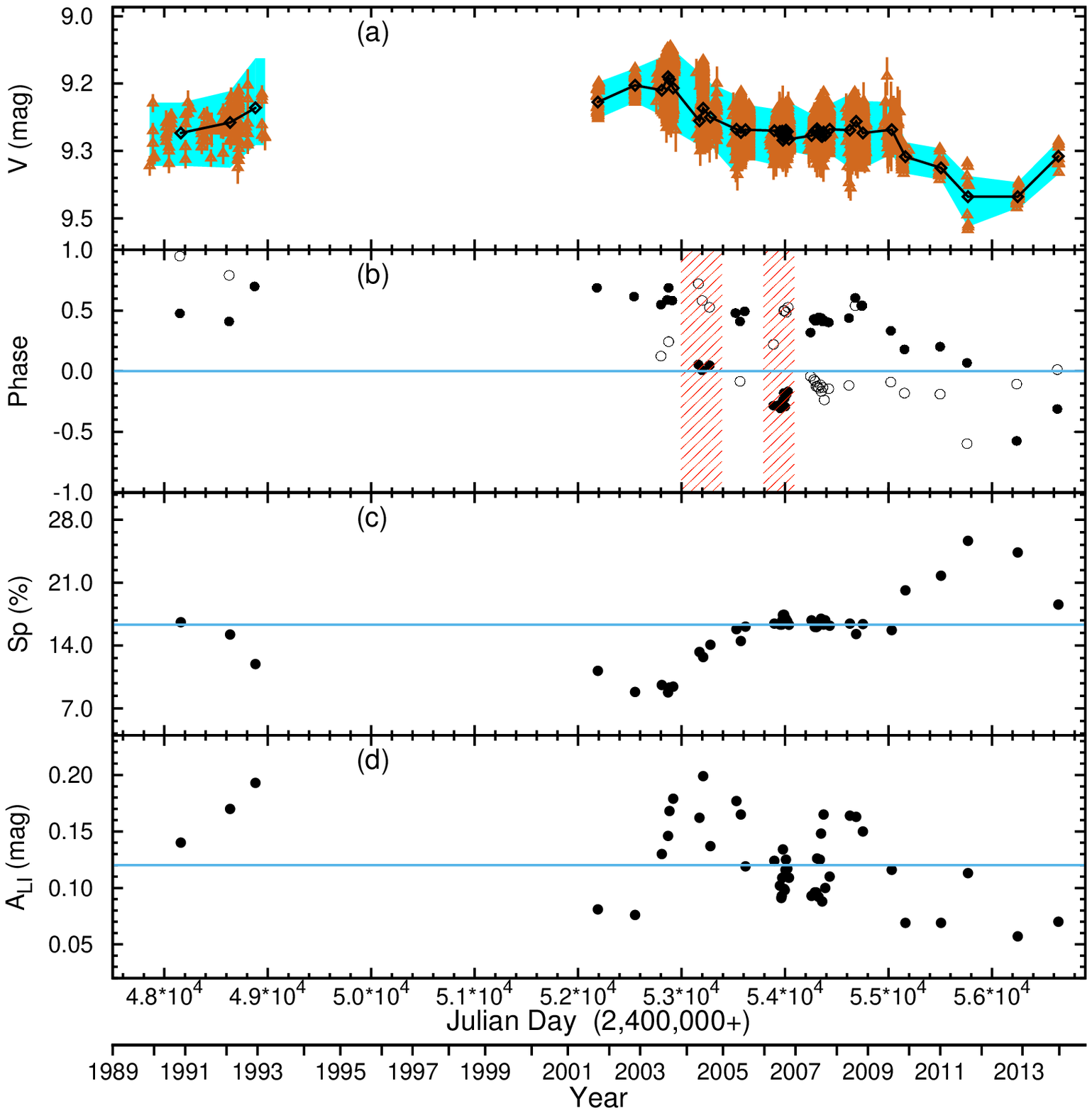}
 \caption{Parameters of LO Peg derived from modelling. From top to bottom -- (a) $V$-band light curve of LO Peg (open triangles) plotted with mean magnitude of each epoch (open diamonds). The shaded regions show the errors in data points. (b) Phases/longitudes of spots recovered from light curve inversion. Filled and open circles show primary and secondary active longitudes, respectively. Vertical shaded regions indicate the time intervals when the possible flip-flop events occur. (c) Recovered surface coverage of cool spots (per cent) on LO Peg. (d) The amplitude of brightness variations in unit of magnitude.} 
 \label{fig_surface-params}
\end{figure*}

%****************************************************************
Fig. \ref{fig_surface-plot} shows the reconstructed temperature inhomogeneity maps of LO Peg, which reveal that the spots have a tendency to concentrate at two longitudes corresponding to two active regions on the stellar surface. The difference between two active longitudes was found to be inconstant. The uncertainty in the positions of the active longitudes on the stellar surface was on average of about 6$^\circ$ (or 0.02 in phase).
The derived stellar parameters  active longitude regions ($\psi_1, \psi_2$), spottedness (Sp), and $V$-band amplitudes ($A_{\rm LI}$) corresponding to each surface brightness map shown in Fig. \ref{fig_surface-plot} are plotted with HJD in Fig. \ref{fig_surface-params}.
The filled and open circles in Fig. \ref{fig_surface-params}(b) show high and low active regions, respectively. The derived parameters are also given in Table \ref{tab_surface-params}.  
 First three surface maps were created with the sparse data of \hipp ~satellite, and to create good quality maps, data of $\sim$2, $\sim$0.5, and $\sim$0.5 yr were used. We get a signature of presence of two equal spot groups $\sim$170$\deg$ apart in first two years (1989 to 1991), whereas the presence of single spot group was indicated with the surface map of the third year (1992). From 2001 to 2013 with ground-based observations and archival data it became possible to create at least one surface map per year.
Using high-cadence data of \wasp ~in 2006 and 2007, we created 12 surface maps (Set-15 to Set-26) in three months of 2006 (July 27th to November 4th) and 10 surface maps (Set-27 to Set-36) in three months of 2007 (July 24th to December 19th). This enables us to make a detailed study on the surface structure of LO Peg.
It appears that LO Peg consists only one spot group during the observations of 2006 August (Set-16 to Set-21). In 2006 September, migration from single spotted surface to double spotted surface was clearly noticeable (Set-22 to Set-26). Both spots are separated by $<$ 115\deg. Two spot groups were also observed during 2007, but the separations of two spot groups were $>$ 125\deg. It was also noticed that the active regions changed their position from 2006 to 2007, which indicates a flip-flop cycle of $\sim$1 yr (see shaded regions on the Fig. \ref{fig_surface-params}b). Similar phenomena were also noticed during the year 2004 and 2005 with an approximately same period. But due to uncertainty on the position of the spot-groups, it was not possible to say whether the flip-flop cyclic behaviour continues in later years.
%*********** fig :9: Surface Imaging Parameters *****************

%****************************************************************
      
The total area of the visible stellar surface covered by spots, known as spottedness (Sp), varies within a range of 8.8--25.7\% (see Fig. \ref{fig_surface-params}c),  with a median value of 16.3\%. From the year 2001 to 2005, it was found to decrease until its minimum in 2003 August, and then return to its median value in 2005. It remained constant for $\sim$4 yr at this value, and then increased to reach its maximum. Further the spottedness of the star has returned to its median value again. The time interval of returning to its median value was approximately same as 4 yr.
 As seen in Fig. \ref{fig_surface-params}(d), the amplitude of the brightness varies within a range 0.06--0.19 mag, with a median value of 0.12 mag.

%%%%%%%%%% SECTION 7 %%%%%%%%%%%%%%%%%%%%%%%%%%%%%%%%%%%%%%%%%%%%%%%%%%%%%%%%%%%%%
\subsection{Coronal and Chromospheric features}
\label{subsec:x-ray}
%*********** fig :10: correlation : X-Ray ~ UV ******************
\begin{figure}
\centering
\includegraphics[width=8cm,angle=-90]{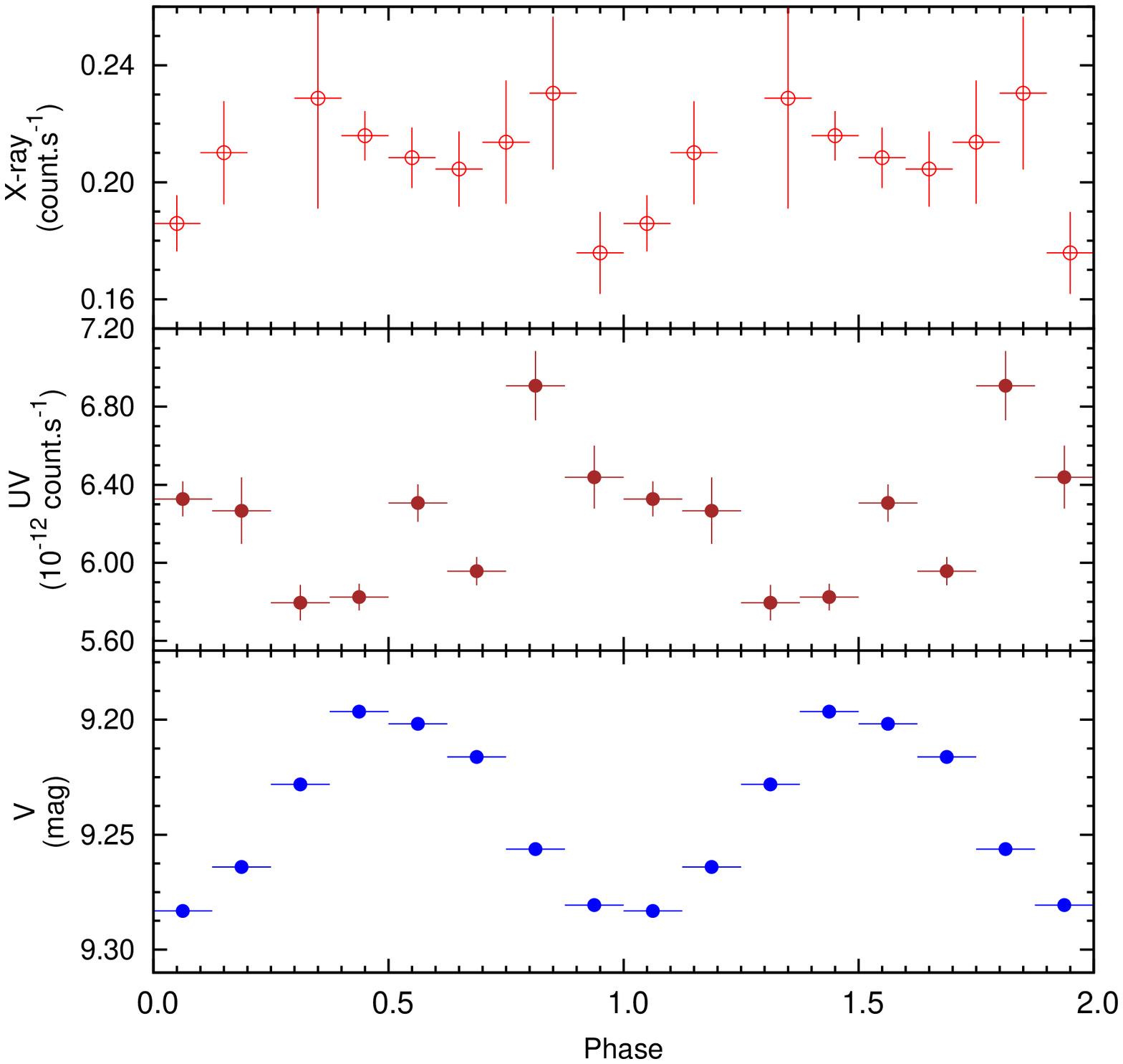}
\caption{From top to bottom the X-ray, UV, and optical folded light curves are shown. Each folded light curve is binned at bin-size 0.1.}
\label{fig_quasi}
\end{figure}

%****************************************************************
The background subtracted X-ray light curves of LO Peg as observed with \swift ~XRT and \rosat ~PSPC instruments are shown in the top panel of Fig. \ref{fig_multiband-lc}. The temporal binning of the X-ray light curves are 100 s. XRT light curves were obtained in energy band 0.3--10.0 keV, whereas \rosat ~light curves were obtained in an energy band 0.3--2.0 keV.  \rosat ~PSPC count rate were converted to \swift ~XRT count rate using {\sc webpimms}\footnote{http://heasarc.nasa.gov/cgi-bin/Tools/w3pimms/w3pimms.pl} where we assumed two temperature components 0.27 and 1.08 keV and 0.2 solar abundances.
To check for the variability, the significance of deviations from the mean count rate were measured using the standard $\chi^2$-test. For our X-ray light curves, derived value of $\chi^2$ is 664 which is very large in comparison to the 190 degrees of freedom ($\chi^2_\nu$ = 3.5). This indicates that LO Peg is essentially variable in X-ray band.

On one occasion (ID: 00037810011) sudden enhancement of X-ray count rates was detected  along with a simultaneous enhancement in UV count rates in each of the UV filters. This enhancement could be due to flaring activity, where flare peak count rates were $\sim$3 times higher than a quiescent level of 0.20 counts s$^{-1}$. The close inspection of the X-ray light curve shows the decay phase of the flare. We could not analyse this flare due to poor statistics. However, the flare duration was found to be 1.2 ks. 
%*********** fig :11: X-ray Spectra *****************************
\begin{figure}
\centering
\includegraphics[width=6cm,angle=-90]
{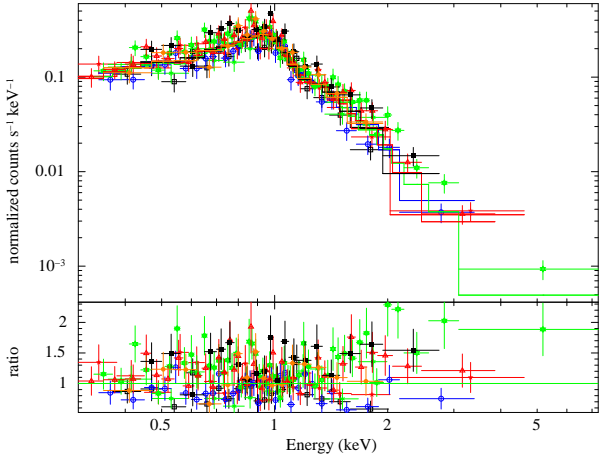}
\caption{X-ray Spectra of LO Peg obtained from \swift ~XRT along with the best fitting {\sc APEC} 2T model (top panel). Different symbols denote different observation IDs. The bottom panel represents the ratio of the observed counts to the counts predicted by best-fitting model.}
\label{fig_x-ray-spectra}
\end{figure}

%****************************************************************

Fig. \ref{fig_quasi} shows the rotationally modulated X-ray, UV (uvm2 filter), and optical ($V$-band) light curves. 
Observations of the year 2008 were used for rotational modulation, where we have removed the flaring feature from X-ray light curve. Optical and X-ray observations are $\sim$100 d apart. It appears that both X-ray and UV light curves were rotationally modulated. X-ray and UV light curves appear to be anti-correlated with $V$-band light curve. The Pearson correlation coefficients between X-ray and $V$-band, and UV and $V$-band light curve were found to be --0.22 and --0.57, respectively.

The \swift ~XRT spectra of the star LO Peg, as shown in Fig. \ref{fig_x-ray-spectra}, were best fitted with two  temperature (2T) astrophysical plasma model  \citep[{\sc apec};][]{smith-01-62}, with variable elemental abundances (Z). The interstellar hydrogen column density (N$_{\rm H}$) was left free to vary. Since all the parameters were found to be constant within a 1$\sigma$ level, we determined the parameters from joint spectral fitting. The two temperatures and corresponding emission measures were 0.28 $\pm$ 0.04 keV and 1.03 $\pm$ 0.05 keV, and 3.1 $\pm$ 0.9 \E{52} cm$^{-3}$ and 4.6 $\pm$ 0.6 \E{52} cm$^{-3}$, respectively. Global abundances were found to be 0.13 $\pm$ 0.02 solar unit (\zsun). The derived value of unabsorbed luminosity is given by $1.4_{-0.4}^{+0.5}$ \E{29} erg s$^{-1}$ cm$^{-2}$.

%%%%%%%%%% SECTION 9 %%%%%%%%%%%%%%%%%%%%%%%%%%%%%%%%%%%%%%%%%%%%%%%%%%%%%%%%%%%%%
\section{Discussion}
\label{sec:discussion}
Using the long-term $V$-band photometry, we have obtained mean seasonal rotation period of 0.4231 $\pm$ 0.0001 d, which is very similar to the previously determined period \citep{Barnes-05-32}. 
For the first time, we have studied long-term variations in LO Peg.
A long-term periodicity with periods of $\sim$2.2 and $\sim$5.98 yr appears to present in the light curve. However, a long-term continuous monitoring is necessary to confirm any long-term periodicity.
The first period is also found to be similar to the latitudinal spot migration period derived from SDR analysis, which could be similar to the 11 yr cycle of the solar butterfly diagram. This type of activity cycle was also observed in similar fast-rotating stars such as  AB Dor \citep{Collier-02-7, Jarvinen-05-4} and LQ Hya \citep{Messina-03-5}.
In the SDR analysis, the decrease in photometric periods within most of the cycles is reminiscent of the sunspot cyclic behaviour, where the latitude of spot-forming region moves towards the equator, i.e., toward progressively faster rotating latitudes along an activity cycle, and spot-groups were  present within $\pm$45\deg ~latitude of LO Peg. This finding indicates that LO Peg has a solar-like SDR pattern.  From spectroscopic analysis, \cite{Piluso-08} also detected the presence of such lower latitude spots. It is interesting to note that the slope of the rotational period on LO Peg varies and therefore SDR amplitude $\Delta P$ (= $ P_{\rm max} - P_{\rm min}$) changes from cycle to cycle. Similar behaviour was also observed in AB Dor \citep{Collier-02-7}, BE Cet, DX Leo, and LQ Hya \citep{Messina-03-5}. This resembles either a wave of excess rotation on a time-scale of the order of decades, or a variation of the width of the latitude band in which spots occur.
LO Peg shows a change in rotation period from 0.43133 to 0.41743 d, which corresponds to $\sim$3 km s$^{-1}$ change in $v$sin$i$, which is nearly 15 times more than AB Dor. We have estimated the differential rotation on LO Peg with $\Delta \Omega/\Omega$ ranging from 0.001--0.03, which is similar to that obtained by \cite{Barnes-05-32}. AB Dor and LQ Hya having very similar spectral class and periodicity showed similar feature with the star LO Peg. Derived values of $\Delta \Omega/\Omega$ for AB Dor \citep{Collier-02-7} and LQ Hya \citep{Berdyugina-02-2} are also similar to that for LO Peg.
During more than two decades of observations only in 2003 September LO Peg was observed both photometrically \citep{Tas-11} and spectroscopically \citep{Piluso-08}. In our SDR analysis, the first point of cycle-VI corresponds to that time interval (see Table \ref{tab_diff-rot} and Fig. \ref{fig_diff-rot}). This time interval being at the starting of the cycle indicates the period corresponds to higher latitude. Therefore, we expect presence of spots on higher latitude on the surface of LO Peg. \cite{Piluso-08} have also found star-spot concentration towards the polar region.

\begin{figure*}
\centering
\includegraphics[width=11cm,angle=-90]
{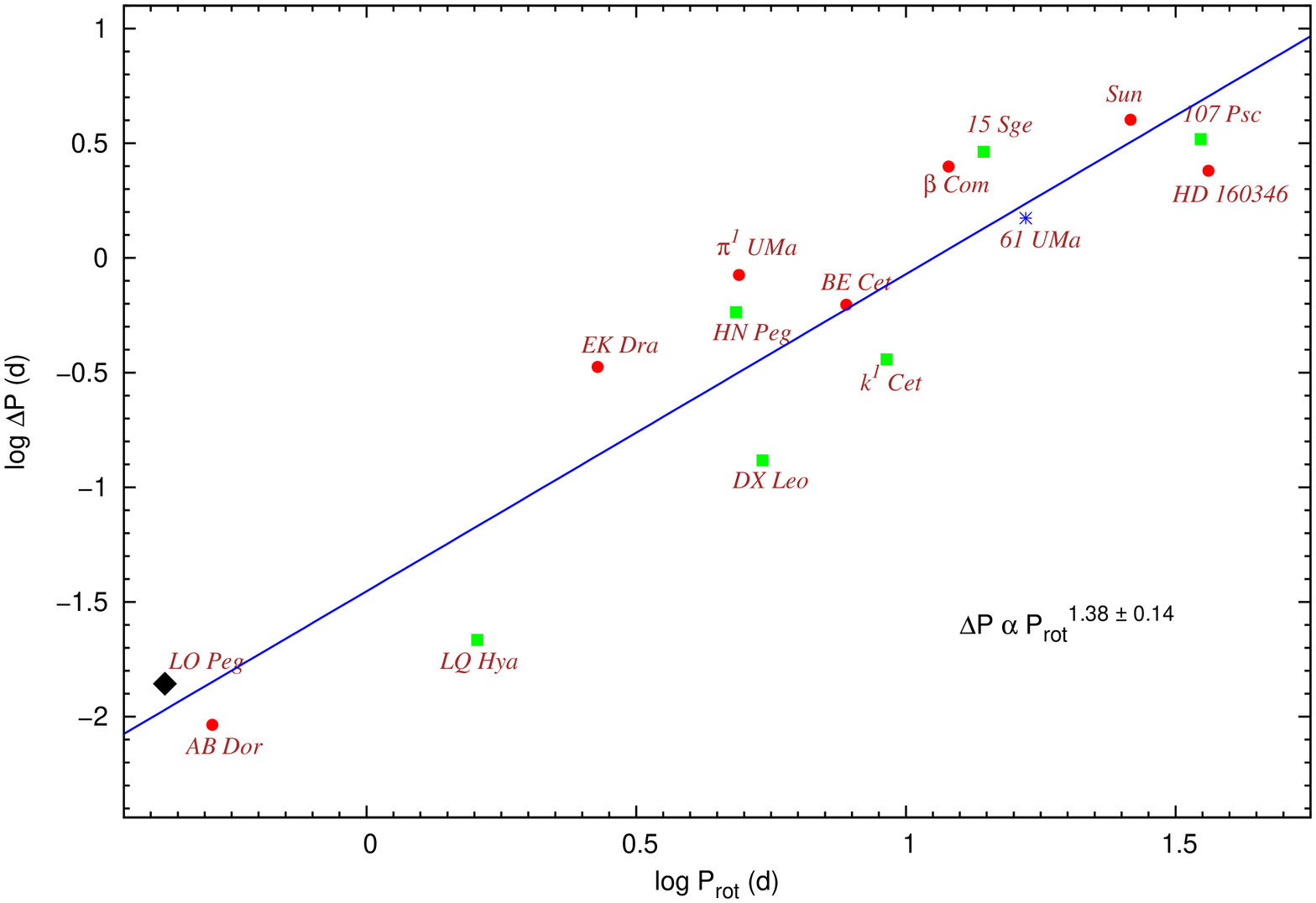}
\caption{The log rotational period variations versus the mean rotational period of stars. Solid circle, solid squares and asterisks denote the stars with solar, anti-solar and hybrid pattern. LO Peg is shown with solid diamond. The continuous line is a power-law fit to the whole sample.}
\label{fig_disc-diff-rot}
\end{figure*}

 A positive correlation between the absolute value of SDR and the stellar rotation period was predicted by dynamo models according to a power law \citep{Kitchatinov-99-3} i.e. $\Delta P~ \alpha~ P^{n}_{\rm rot}$; where $\Delta P$ is the SDR amplitude, $P_{\rm rot}$ is the rotational period and $n$ is the power index. \cite{Kitchatinov-99-3} found that n varies with both rotation rate and with spectral type. This power-law dependence is confirmed by observational data \citep{Hall-91-15, Henry-95-1}, although the observational and theoretical values of $n$ differ \citep[see][]{Messina-03-5}. 
Fig. \ref{fig_disc-diff-rot} shows the plot between $\Delta P$ and $P_{\rm rot}$ of LO Peg with other 14 stars with known activity cycles and SDR \citep{Messina-03-5, Collier-02-7, Donahue-96-11, Gray-97-9}. We found LO Peg (solid diamond) follows the same trend with the nearest candidate AB Dor. Including LO Peg, we derive the relation $\Delta P~ \alpha~ P^{1.4 \pm 0.1}_{\rm rot}$, which is  very similar to the relations derived from other observational evidences such as $n$ = 1.4 $\pm$ 0.5 \citep{Messina-03-5}, $n$ = 1.30 \citep{Donahue-96-11}, $n$ = 1.15--1.30 \citep{Rudiger-98-3}. This indicates the disagreement between the observational ($n$ = 1.1--1.4) and the theoretical ($n$ $>$ 2) values of power-law index \citep{Kitchatinov-99-3}. 

Present surface imaging indicates that, in most of the cases the spots on LO Peg were concentrated in two groups separated by less than 180\deg ~along the longitude. Indication of the flip-flop effect in LO Peg is quite similar to that observed by  \cite{Korhonen-02-2} and  \cite{Jarvinen-05-2}. 
The flip-flop phenomenon has been noticed for the first  time by \cite{Jetsu-91-3} in the giant star FK Com. Later it was found to be cyclic in RS CVn and FK Com-type stars, as well as in some young solar analogues \citep[e.g. ][]{Korhonen-02-2}. After its discovery in cool stars, the flip-flop phenomena have also been reported in the Sun \citep{Berdyugina-03-5}. This phenomenon is well explained by the dynamo based solution  where  a non-axisymmetric dynamo component, giving rise to two permanent active longitudes 180$^\circ $ apart, is needed together with an oscillating axisymmetric magnetic field \citep{Elstner-05-2, Korhonen-05-1}. \cite{Fluri-04-1} suggest  another possibility with a combination of stationary axisymmetric and varying non-axisymmetric components. 
It also appears that, the flip-flop cycle is approximately one third of the latitudinal spot migration cycle.
%*********** fig :12: Discussion: Diff-rotation *****************

%****************************************************************

Modelling of LO Peg reveals that the stellar surface is spotted up to 25.7\%, which is very similar to that found in  K-type stars XX Tri \citep{Savanov-14-2}, V1147 Tau \citep{Patel-13-20}, LQ Hya, and MS Ser  \citep{Alekseev-03-5}.
We did not see any relation between spottedness and cyclic behaviour or the rotational period. However, from the year 2005 to 2009 spottedness variation is found to be almost constant (shown in third panel of Fig \ref{fig_surface-params}). At the same time duration, SDR analysis also indicates that the seasonal rotational period and hence the latitude of the spot groups also remains constant (see cycle-VII in second panel of Fig. \ref{fig_diff-rot}). This suggests that the magnetic activities remains constant within that period of time. In all other cycles the seasonal rotational period and hence latitudinal spot groups follows solar-like butterfly pattern with a $\sim$2.7 yr period. Whereas, spottedness variation  does not show any periodic modulation.
The observations of Set-8 (2003 September 11--October 30) are quasi-simultaneous with the spectroscopic observation of \cite{Piluso-08}. During this period our analysis shows single spot at phase 210\deg. Whereas Doppler imaging study of \cite{Piluso-08} shows a signature of low-latitude spot at phase $\sim$0.7. Correcting for the difference in ephemeris, we get the corresponding star-spot longitude to be $\sim$226$\deg$, which is almost similar to our derived value. The longer time span used in the generation of surface map may cause the difference between the two longitude positions.    
 The brightness variability amplitudes of LO Peg were found to vary from 0.06 to 0.19 mag. This value is very similar to variability amplitude of other K-type stars such as V1147 Tau \citep{Patel-13-20}, LQ Hya \citep{Berdyugina-02-2}, AB Dor \citep{Jarvinen-05-4}, and MS Ser \citep{Alekseev-03-5}. \cite{Savanov-14-2} detected the variability amplitude up to  0.8 mag in late-type star ASAS 063656-0521.0. 

\begin{figure*}
\centering
\subfigure[]{\includegraphics[width=6cm,angle=-90]{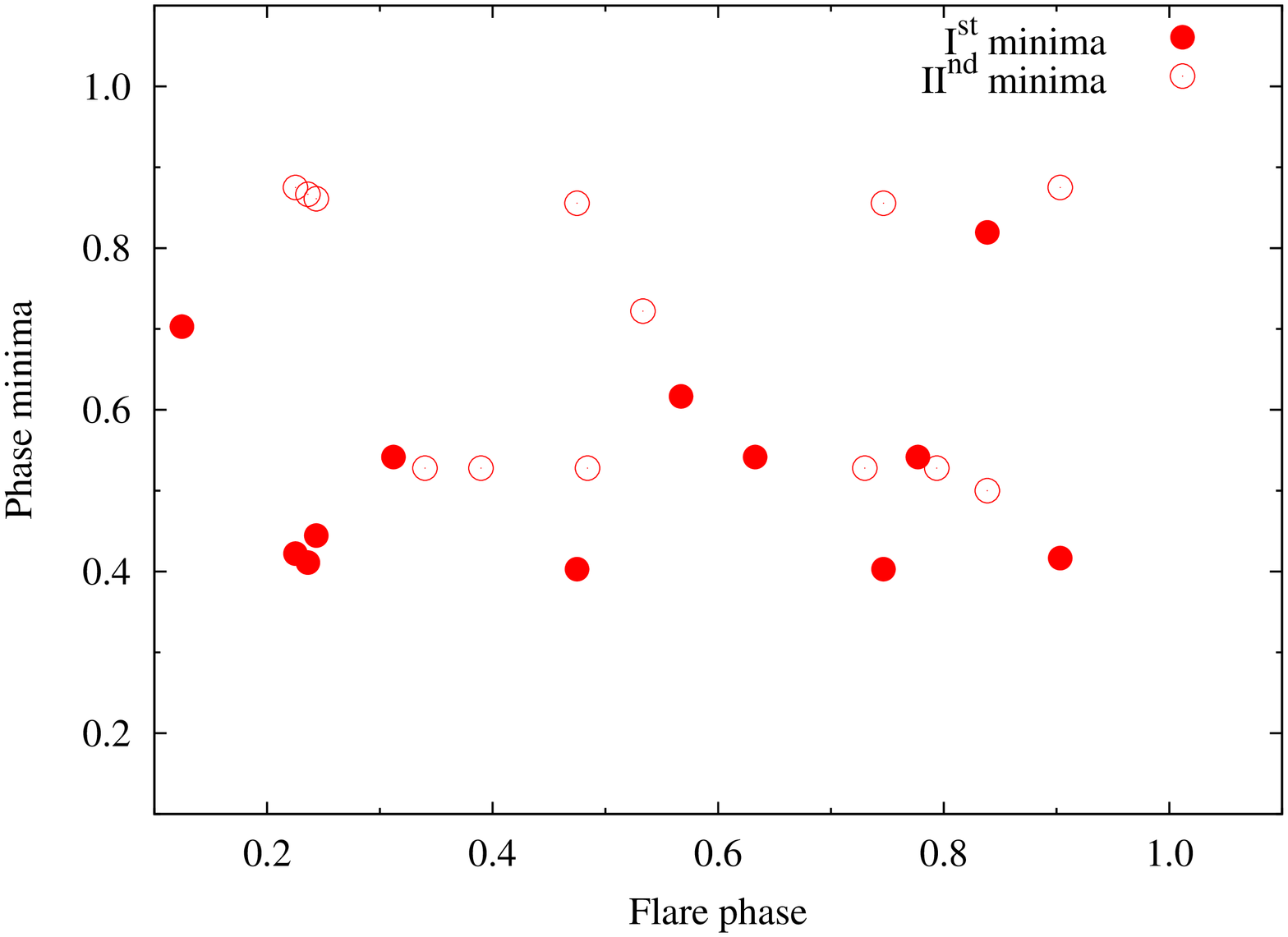}}
\subfigure[]{\includegraphics[width=6cm,angle=-90]{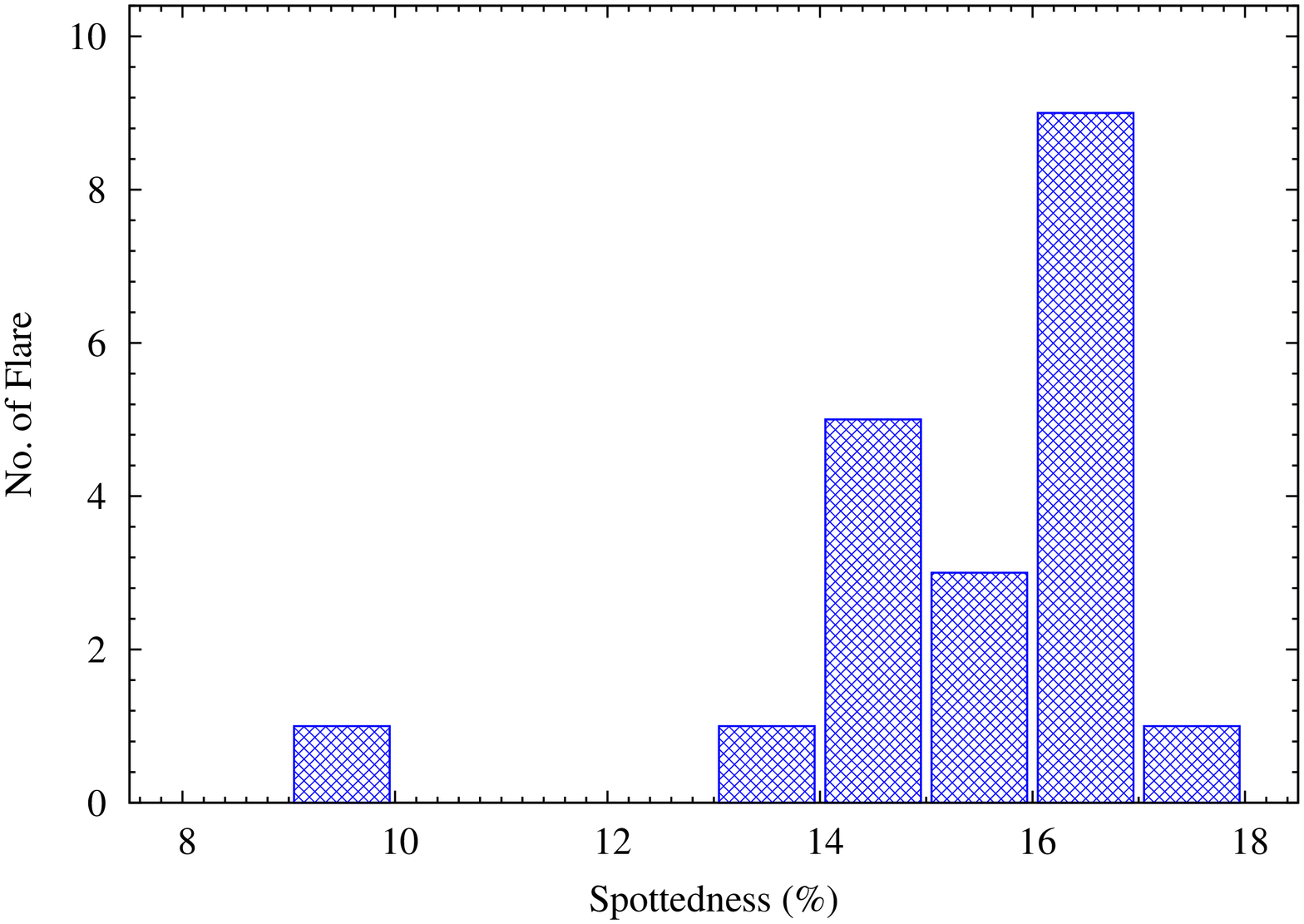}}
\caption{(a) Phase minima is plotted as a function of flare peak phase. (b) Observed distribution of detected flares with stellar spottedness.}
\label{fig_disc-sp-fl}
\end{figure*}

In our multiband photometric study, we detected a total of 20 optical flares with a frequency of $\sim$1 flare per two days.  There are very few detailed studies of optical flares on UFRs due to constraints in their detection limit, detection timing, and very less flare frequency. Recent studies on optical flares done with ground-based observatories on DV Psc \citep{Pi-14-7} and CU Cnc \citep{Qian-12-18} show flare frequency of $\sim$2 flares per day and $\sim$1 flare per day, respectively, which is similar to that
for LO Peg. However, several other studies done with \kepler ~satellite \citep[][]{Hawley-14,Lurie-15-1} on M dwarfs reveal that flare frequency varies over a wide range of $\sim$1 flare per month to $\sim$10 flares per day.
In our study, we have also detected an X-ray flare simultaneously observed with the UV-band. Derived value of X-ray flare frequency is $\sim$3 flares per day. This suggests that LO Peg shows more activity in X-rays than in optical bands. 
 Multiwavelength simultaneous studies of another UFR AB Dor also shows a similar  behaviour \citep{Lalitha-13-1}. This implies that the corona is more active in comparison to photosphere in these stars.
 Most of the flares are $\sim$1 h long with a minimum and maximum flare duration of $\sim$12 min and $\sim$3.4 h. Flare observed on SV Cam \citep{Patkos-81-2}, XY UMa \citep{Zeilik-82-4}, DK CVn \citep{Dal-12-6}, FR Cnc \citep{Golovin-12-2},  AB Dor \citep{Lalitha-13-1}, and DV Psc \citep{Pi-14-7} studied in optical bands also lie in the same range. Several flares observed on M dwarfs by \cite{Hawley-14} show similar feature but with a flare duration of $\sim$2 min. The derived flare amplitudes on LO Peg were found to be higher in shorter wavelengths than that in longer wavelengths (see two representative flares in Fig. \ref{fig_flare-multi}). Similar feature was also found in multiwavelength studies of the flare on FR Cnc \citep{Golovin-12-2}. Although most of the flares occurred on LO Peg show the usual fast rise and slow decay, there were a few flares that show the reverse phenomena. This was also previously observed on K-type star V711 Tau \citep{Zhang-90-33}, which may be the result of complex flaring activity at the rise phase of the flare which could not be resolved due to instrumental limitations. \cite{Davenport-14-1} show the existence of complex flares with high-cadence \kepler ~data which can be explained by a superposition of multiple flares.     
 Flare energies derived for LO Peg lies in the range of $\sim$\Pten{31-34} erg. Seventeen out of twenty flares having energy less than \Pten{33} erg, signifies more energetic flares are less in number. With \kepler ~data \cite{Hawley-14} also detected most of the flares on M dwarf GJ 1243 having energy of the order of \Pten{31} erg. 
One flare (F13) is found to have total energy more than \Pten{34} erg, therefore, this flare can be classified as a Superflare \citep[see][]{Candelaresi-14-1}. The derived energy of this flare was $\sim$10.5 times more than the next largest flare and  668 times more than the weakest flare observed on LO Peg. During the flare F13, the $V$-band magnitude increases up to 0.42 mag, similar enhancement in $V$-band mag also noticed in FR Cnc \citep{Golovin-12-2}. 
We inspected the list of flares for further evidence of a correlation between flare timing and orientation of the dominant spot group. The phase minima as a function of flare phase is plotted in Fig. \ref{fig_disc-sp-fl}(a), where we did not find any correlation. This finding is also consistent with the flare study of \cite{Hunt-Walker-12-1} and \cite{Roettenbacher-13-1}. From this result we conclude that most of the flares on LO Peg may not originate in the strongest spot group, but rather come from small spot structures or polar spots.
 In order to check whether the spottedness on the stellar surface is related to occurrence rate of flare, we plotted the distribution of detected flares in each percentage binning of spottedness shown in Fig. \ref{fig_disc-sp-fl}(b). Most of the flares detected on LO Peg are found to occur within a spottedness range of 13--18\%, with a highest number of 9 flares occurred at a spottedness range of 16--17\%.
%*********** fig :14: Discussion: Spottedness vs. #of Flare *****

%****************************************************************

The coronal parameters derived in this study are very similar to that derived by \cite{Pandey-05-8} using \rosat ~data. The corona of LO Peg consists of two temperature, which is similar to few UFRs such as Speedy Mic, YY Gem and HK Aqr \citep{Singh-99-22}, whereas it differs from others UFRs such as AB Dor, HD 283572, and EK Dra \citep{Gudel-01-4, Scelsi-05-2}, which consists of three temperature corona. From the present analysis, light curve in X-ray and UV band were found to be anti-correlated with optical $V$-band. This feature was also noticed in similar type of stars such as  HR 1099, $\sigma$ Gem, V1147 Tau, and AB Dor \citep{Agrawal-88-1, Lalitha-13-1, Patel-13-20} which indicates the presence of high chromospheric and coronal activity in the spotted region.

%%%%%%%%%% SUMMARY %%%%%%%%%%%%%%%%%%%%%%%%%%%%%%%%%%%%%%%%%%%%%%%%%%%%%%
\section{Summary}
\label{sec:summary}
In this study, with $\sim$24 yr long photometric observations from different worldwide telescopes, and X-ray and UV observations obtained with \swift ~satellite we have investigated the properties of a UFR LO Peg. The results of this study are summarized as below.

\begin{itemize}

\item{The rotational period of LO Peg steadily decreases along the activity cycle, jumping back to higher values at the beginning of the next cycle with a cycle of 2.7 $\pm$ 0.1 yr, indicating a solar-like SDR pattern on LO Peg.}

\item{ We have detected  20 optical flares, where the most energetic flare has energy of $10^{34.2}$ erg whereas the least energetic flare has energy of $10^{30.9}$ erg with flare duration range of 12--202 min.}

\item{Our inversion of phased light curves show the surface coverage of cool spots are in the range of $\sim 9 - 26$ per cent. Evidence of flip-flop cycle of $\sim$1 yr is also found.}

\item{ Corona of LO Peg consist of  two temperatures of $\sim$3 MK and $\sim$12 MK. Quasi-simultaneous observations in X-ray, UV, and optical $UBVR$ bands show
a signature of high X-ray and UV activities in the direction of spotted regions.}
\end{itemize}

%%%%%%%%%% ACKNOWLEDGEMENTS %%%%%%%%%%%%%%%%%%%%%%%%%%%%%%%%%%%%
\section*{Acknowledgments}
We thank the referee for his/her comments and suggestions that helped to considerably improve the manuscript. This research has been done under the Indo--Russian DST-RFBR project reference INT/RUS/RFBR/P-167 (for India) and Grant RFBR Ind\_a 14-02-92694 (for Russia). SJ acknowledges the grant received under the Indo--Russian DST-RFBR project reference INT/RUS/RFBR/P-118. We acknowledge NASA Exoplanet Archive, All Sky Automated Survey archive, \hipp ~archive, \swift ~archive, and  different telescope facilities we used to carry out our research.

%%%%%%%%%%% REFERENCES %%%%%%%%%%%%%%%%%%%%%%%%%%%%%%%%%%%%%%%%%%
 \bibliography{SK_collections.bib}{}

%%%%%%%%%%%% FIGURES %%%%%%%%%%%%%%%%%%%%%%%%%%%%%%%%%%%%%%%%%%%%
%%%%%%%%%%%%%%%% END %%%%%%%%%%%%%%%%%%%%%%%%%%%%%%%%%%%%%%%%%
\end{document}